\documentclass[journal, a4paper]{IEEEtran}

\IEEEoverridecommandlockouts

\usepackage{algorithm,algorithmic}
\usepackage{amsmath,amssymb,mathrsfs}
\usepackage{array}
\newcolumntype{I}{!{\vrule width 2pt}}
\newlength\savedwidth

\newlength\savewidth

\usepackage{balance}
\usepackage{booktabs}
\usepackage{cite}
\usepackage{epsfig}
\usepackage{enumitem}
\usepackage{easyReview}
\usepackage{fancyhdr}
\usepackage{float}
\usepackage[nomain,acronym,automake,toc,nopostdot]{glossaries}
\usepackage{graphicx,color}
\usepackage{hyperref}
\usepackage{mathtools}
\usepackage{multirow}
\usepackage{psfrag}
\usepackage{xcolor}
\usepackage{stfloats}
\usepackage{subfigure}
\usepackage{url}

\newtheorem{thm}{Theorem}
\newtheorem{cor}{Corollary}[thm]

\newtheorem{defn}{Definition}

\newcommand{\mat}[1]{\mbox{\boldmath $#1$}}
\newcommand{\algref}[1]{\textbf{Algorithm \ref{#1}}}

\newcommand{\figref}[1]{Fig. \ref{#1}}
\newcommand{\secref}[1]{Section \ref{#1}}
\newcommand{\appref}[1]{Appendix \ref{#1}}
\newcommand{\tabref}[1]{TABLE \ref{#1}}
\renewcommand{\eqref}[1]{(\ref{#1})}

\definecolor{sblue}{RGB}{0,51,120}
\definecolor{sred}{RGB}{139,0,139}
\definecolor{sg}{RGB}{46,139,87}

\DeclareMathOperator*{\argmin}{arg\,min}

\makeglossaries
\newacronym[description=Additive white Gaussian noise]{awgn}{AWGN}{additive white Gaussian noise}
\newacronym[description=Approximate message passing]{amp}{AMP}{approximate message passing}
\newacronym[description={\em a posteriori} probability]{app}{APP}{{\em a posteriori} probability}
\newacronym[description=Base station]{bs}{BS}{base station}
\newacronym[description=Base station sleeping ]{bss}{BSS}{base station sleeping }
\newacronym[description=Belief propagation]{bp}{BP}{belief propagation}
\newacronym[description=Binary phase shift keying]{bpsk}{BPSK}{binary phase shift keying}
\newacronym[description=Bit error rate]{ber}{BER}{bit-error-rate}
\newacronym[description=Block error rate]{bler}{BLER}{block error rate}
\newacronym[description=Central limit theorem]{clt}{CLT}{central limit theorem}
\newacronym[description=Channel state information ]{csi}{CSI}{channel state information }
\newacronym[description=Closest vector problem]{cvp}{CVP}{closest vector problem}
\newacronym[description=Code division multiple access]{cdma}{CDMA}{code division multiple access}
\newacronym[description=Distributed linear data fusion]{dldf}{DLDF}{distributed linear data fusion}
\newacronym[description=European Cooperation in Science and Technology]{cost}{COST}{Cooperation in Science and Technology}
\newacronym[description=Coordinated multi-point ]{CoMP}{CoMP}{coordinated multi-point }
\newacronym[description=Correlation-based stochastic model]{cbsm}{CBSM}{correlation-based stochastic model}
\newacronym[description=Cumulative distribution function]{cdf}{CDF}{cumulative distribution function}
\newacronym[description=Degrees of freedom]{dof}{DoF}{degrees of freedom}
\newacronym[description=Element-based lattice reduction]{elr}{ELR}{element-based lattice reduction}
\newacronym[description=Extremely-large aperture array]{elaa}{ELAA}{extremely-large aperture array}
\newacronym[description=Fifth-generation]{5g}{5G}{fifth-generation}
\newacronym[description=Fixed-complexity sphere decoder]{fcsd}{FCSD}{fixed-complexity sphere decoder}
\newacronym[description=Forward error corrrection]{fec}{FEC}{forward error correction}
\newacronym[description=Free space path loss]{fspl}{FSPL}{free space path loss}
\newacronym[description=Global system for mobile communication]{gsm}{GSM}{global system for mobile communication}
\newacronym[description=Geometry-based stochastic model]{gbsm}{GBSM}{geometry-based stochastic model}
\newacronym[description=Hermite-Korkin-Zolotarev]{hkz}{HKZ}{Hermite-Korkin-Zolotarev}
\newacronym[description=Independent and identically distributed]{iid}{i.i.d.}{independent and identically distributed}
\newacronym[description=Integer least-squares]{ils}{ILS}{integer least-squares}
\newacronym[description=International Telecommunication Union Radiocommunication Sector ]{itu-r}{ITU-R}{International Telecommunication Union Radiocommunication Sector}
\newacronym[description=Kullback–Leibler]{kl}{KL}{Kullback–Leibler}
\newacronym[description=Large system behaviour]{lsb}{LSB}{large system behaviour}
\newacronym[description=Lattice reduction]{lr}{LR}{lattice reduction}
\newacronym[description=Lenstra-Lenstra-Lov\'{a}sz]{lll}{LLL}{Lenstra-Lenstra-Lov\'{a}sz}
\newacronym[description=Likelihood ascent search]{las}{LAS}{likelihood ascent search}
\newacronym[description=Line-of-slight]{los}{LoS}{line-of-slight}
\newacronym[description=List sphere decoder]{lsd}{LSD}{list sphere decoder}
\newacronym[description=Linear minimum mean square error]{lmmse}{LMMSE}{linear minimum mean square error}
\newacronym[description=Log-likelihood ratio]{llr}{LLR}{log-likelihood ratio}
\newacronym[description=Long-term evolution ]{lte}{LTE}{long-term evolution}
\newacronym[description=Low density parity check]{ldpc}{LDPC}{low density parity check}
\newacronym[description=Massive machine type communications]{mmtc}{mMTC}{massive machine type communications}
\newacronym[description=Maximum {\em a posteriori}]{map}{MAP}{maximum {\em a posteriori}}
\newacronym[description=Maximum-likelihood estimation]{mle}{MLE}{maximum-likelihood estimation}
\newacronym[description=Maximum-likelihood sequence detection]{mlsd}{MLSD}{maximum-likelihood sequence detection}
\newacronym[description=Multiple-input multiple-output]{mimo}{MIMO}{multiple-input multiple-output}
\newacronym[description=Multipath components]{mpc}{MPC}{multipath components}
\newacronym[description=massive multiple-input multiple-output]{mmimo}{mMIMO}{massive multiple-input multiple-output}
\newacronym[description=Matched filter]{mf}{MF}{matched filter}
\newacronym[description=Maximum ratio combining]{mrc}{MRC}{maximum ratio combining}
\newacronym[description=Mean square error]{mse}{MSE}{mean square error}
\newacronym[description=Minimum mean square error]{mmse}{MMSE}{minimum mean square error}
\newacronym[description=Mobile and wireless communications Enablers for the Twenty-twenty Information ]{metis}{METIS}{Mobile and wireless communications Enablers for the Twenty-twenty Information}
\newacronym[description=Non-line-of-sight]{nlos}{NLoS}{non-LoS}
\newacronym[description=One dimensional]{1d}{1-D}{one dimensional}
\newacronym[description=Orthogonality defect]{od}{OD}{orthogonality defect}
\newacronym[description=Pairwise error probability]{pep}{PEP}{pairwise error probability}
\newacronym[description=Parallel interference cancellation]{pic}{PIC}{parallel interference cancellation}
\newacronym[description=Probabilistic data association]{pda}{PDA}{probabilistic data association}
\newacronym[description=Probability distribution function]{pdf}{PDF}{probability density function}
\newacronym[description=Probability mass function]{pmf}{PMF}{probability mass function}
\newacronym[description=Quadrature amplitude modulation]{qam}{QAM}{quadrature amplitude modulation}
\newacronym[description=Quadrature phase shift keying]{qpsk}{QPSK}{quadrature phase shift keying}
\newacronym[description=Receiver-side channel state information]{rcsi}{R-CSI}{receiver-side channel state information}
\newacronym[description=Received signal strength]{rss}{RSS}{received signal strength}
\newacronym[description=Semidefinite relaxation]{sdr}{SDR}{semidefinite relaxation}
\newacronym[description=Seysen's algorithm]{sa}{SA}{Seysen's algorithm}
\newacronym[description=Signal-to-interference-plus-noise ratio]{sinr}{SINR}{signal-to-interference-plus-noise ratio}
\newacronym[description=Signal to interference ratio]{sir}{SIR}{signal to interference ratio }
\newacronym[description=Signal-to-noise ratio]{snr}{SNR}{signal-to-noise ratio}
\newacronym[description=Single antenna interference cancellation]{saic}{SAIC}{single antenna interference cancellation}
\newacronym[description=Single input single output]{siso}{SISO}{single input single output}
\newacronym[description=Singular value decomposition ]{svd}{SVD}{singular value decomposition }
\newacronym[description=Sixth-generation mobile networks]{6g}{6G}{sixth-generation}
\newacronym[description=Sphere decoder]{sd}{SD}{sphere decoder}
\newacronym[description=Space-time codes]{stc}{STC}{space-time codes}
\newacronym[description=State-of-the-art]{sota}{SoTA}{state-of-the-art}
\newacronym[description=Successive interference cancellation]{sic}{SIC}{succesive interference cancellation}
\newacronym[description=Symbol error rate]{ser}{SER}{symbol-error-rate}
\newacronym[description=Tabu search]{ts}{TS}{tabu search}
\newacronym[description=Three-dimensional]{3d}{3-D}{three-dimensional}
\newacronym[description=The 3rd Generation Partnership Project]{3gpp}{3GPP}{the 3rd Generation Partnership Project}
\newacronym[description=Two-dimensional]{2d}{2-D}{two-dimensional}
\newacronym[description=Uniform linear array]{ula}{ULA}{uniform linear array}
\newacronym[description=Urban micro]{umi}{UMi}{urban micro}
\newacronym[description=User equipment]{ue}{UE}{user equipment}
\newacronym[description=User terminal]{ut}{UT}{user terminal}
\newacronym[description=Vector error rate]{ver}{VER}{vector error rate}
\newacronym[description=Vertical Bell Labs layered space-time]{vblast}{V-BLAST}{vertical Bell Labs layered space-time}
\newacronym[description=Visibility region]{vr}{VR}{visibility region}
\newacronym[description=Widely linear]{wl}{WL}{widely linear}
\newacronym[description=Widely linear zero forcing]{wlzf}{WLZF}{widely linear zero forcing}
\newacronym[description=Wide-sense stationary uncorrelated scattering]{wssus}{WSSUS}{wide-sense stationary uncorrelated scattering}
\newacronym[description=Wireless World Initiative New Ratio]{winner}{WINNER}{Wireless World Initiative New Ratio}
\newacronym[description=Zero forcing]{zf}{ZF}{zero forcing}
\newacronym[description=Zero mean complex circularly symmetric]{zmccs}{ZMCCS}{zero mean complex circularly symmetric}
\newacronym[description=Access Point]{ap}{AP}{access point}

\ifCLASSINFOpdf
\else
\fi

\begin{document}
\title{A Spatially Non-stationary Fading Channel Model for Simulation and (Semi-) Analytical Study of ELAA-MIMO}
\author{Jiuyu Liu, Yi Ma, and Rahim Tafazolli
	\thanks{Jiuyu Liu, Yi Ma and Rahim Tafazolli are with the 5GIC and 6GIC, Institute for Communication Systems (ICS), University of Surrey, Guildford, United Kingdom, GU2 7XH, e-mail: (jiuyu.liu, y.ma, r.tafazolli)@surrey.ac.uk. ({\em Corresponding author: Yi Ma})}
	\thanks{This work has been partially presented in GLOBECOM'2021, Madrid \cite{Liu2021}.}	
}

\markboth{}%
{Shell \MakeLowercase{\textit{et al.}}: Bare Demo of IEEEtran.cls for IEEE Journals}
\maketitle
\begin{abstract}
	In this paper, a novel spatially non-stationary fading channel model is proposed for multiple-input multiple-output (MIMO) system with extremely-large aperture service-array (ELAA). 
	The proposed model incorporates three key factors which cause the channel spatial non-stationarity: {\em 1)} link-wise path-loss; {\em 2)} shadowing effect; {\em 3)} line-of-sight (LoS)/non-LoS state. 
	With appropriate parameter configurations, the proposed model can be used to generate computer-simulated channel data that matches the published measurement data from practical ELAA-MIMO channels. 
	Given such appealing results, the proposed fading channel model is employed to study the cumulative distribution function (CDF) of ELAA-MIMO channel capacity. 
	For all of our studied scenarios, it is unveiled that the ELAA-MIMO channel capacity obeys the skew normal distribution.
	Moreover, the channel capacity is also found close to the Gaussian or Weibull distribution, depending on users' geo-location and distribution.
	More specifically, for single-user equivalent scenarios or multiuser scenarios with short user-to-ELAA distances (e.g., $1$ m), the channel capacity is close to the Gaussian distribution; and for others, it is close to the Weibull distribution.
	Finally, the proposed channel model is also employed to study the impact of channel spatial non-stationarity on linear MIMO receivers through computer simulations.
	The proposed fading channel model is available at \url{https://github.com/ELAA-MIMO/non-stationary-fading-channel-model}.
\end{abstract}


\IEEEpeerreviewmaketitle

\section{Introduction}\label{sec1}
\IEEEPARstart{C}{onsider} link-level simulation of narrowband wireless multiple-input multiple-output (MIMO) communications in fading channels.
Wireless MIMO channel is mathematically described in form of a transition matrix $\mathbf{H}\in\mathbb{C}^{M\times N}$, which relates the input $\mathbf{s}\in\mathbb{C}^{N}$ to the output $\mathbf{y}\in\mathbb{C}^{M}$ as
\begin{equation}\label{eqn01}
	\mathbf{y}=\mathbf{H}\mathbf{s}+\mathbf{v},
\end{equation}
where $\mathbf{v}\in\mathbb{C}^{M}$ denotes the thermal noise. 
For each Monte Carlo trial, $\mathbf{H}$ is randomly generated according to a certain distribution. 

Conventionally, the wireless MIMO channel is assumed to be \gls{wssus}\cite{Hochwald2004}.
When each element of $\mathbf{H}$ (i.e., $H_{m,n}$) is independently generated with identical probability density function (p.d.f.) denoted by $f_H(x)$, i.e., \footnote{As reminded by a reviewer, the WSSUS assumption does not imply that each channel element can be independently generated. The correlation between different channel elements can still exist.}
\begin{subequations}\label{eqn02}
	\begin{align}
		f_{H_{m,n}}(x)&=f_{H}(x), ~_{\forall m,n};\label{eqn02a}\\
		f_\mathbf{H}(x_{1,1},\cdots,x_{M,N})&=\prod_{m,n}f_{H_{m,n}}(x),\label{eqn02b}
	\end{align}
\end{subequations}
where $f_\mathbf{H}(x_{1,1},\cdots,x_{M,N})$ denotes the joint p.d.f. of all the elements of $\mathbf{H}$. 
In academic publications and industrial technical reports, the most frequently adopted channel p.d.f. is (see \cite{fowc}):
\begin{equation}\label{eqn03}
	f_H(x)=\mathcal{CN}(x; \mu, \sigma^2);
\end{equation}
$\mu$ and $\sigma^2$ denote the mean and variance, respectively, with the following relationship:
\begin{equation}\label{eqn04}
	\mu^*\mu+\sigma^2=\frac{\rho}{d^{\alpha}},
\end{equation}
where  $[\cdot]^*$ stands for the conjugate, $d$ for the Euclidean distance of a pair of transmit-antenna and receive-antenna, $\alpha$ for the path-loss exponent, and $\rho$ for the path-loss coefficient incorporating the system loss, the transmit power, the transmit and receive antenna-gains, etc. (see \cite{Du2010}).
The amplitude of the random variable $(H)$ is well known to obey either the Rice distribution ($\mu\neq0$) or Rayleigh distribution ($\mu=0$) (see \cite{2009wireless}).
It is perhaps worth noting that a more generalized model of fading channels is the Nakagami-{\em m} distribution, which is however not the topic of interest in this work.

\subsection{Motivation of This Work}
Conventional MIMO fading channel models (such as \eqref{eqn02}) are suitable for wireless communications in the far-field region.
However, future MIMO technology requires service antenna-array to have an extremely large aperture (i.e., ELAA-MIMO),
where \glspl{ut} are usually located in the near-field region of the ELAA \cite{BJORNSON20193}.
In this scenario, empirical evidence suggests that the \gls{rss} of ELAA channel can vary widely from link to link, e.g., \cite{Harris2016,7063445,Carton2016}.
To capture this physical characteristic, the plane-wave model, common in conventional MIMO systems, should be replaced by the spherical-wave model \cite{7414041}.
Moreover, transmitter-and-receiver antenna pairs can experience different \gls{los}/\gls{nlos} conditions as well as different shadowing effects \cite{Wang2018,8866736,metis}.
All these factors contribute to the spatial non-stationarity of the ELAA channel, which \eqref{eqn02} fails to encapsulate.
Moreover, our state-of-the-art analysis in Section \ref{1b} shows that current MIMO fading channel models are too simple to capture key features of the ELAA channel spatial non-stationarities, and other channel models are too complex for both computer simulations and analytical study. 
To facilitate the link-level study of ELAA-MIMO systems, a spatially non-stationary fading channel model is required.

\subsection{State-of-the-Art Analysis}\label{1b}
\begin{table*}[t]
	\centering
	\renewcommand{\arraystretch}{1.5}
	\caption{A brief review of current massive-MIMO channel models (see details in \cite{Wang2018})}
	\vspace{-0em}
	\label{tabR1}
	\resizebox{1\textwidth}{!}{%
		\begin{tabular}{c>{\raggedright\arraybackslash}p{3.7cm}>{\raggedright\arraybackslash}p{5.6cm}>{\raggedright\arraybackslash}p{3.2cm}}
			\toprule[1.5pt]
			\textbf{Model} & \textbf{Structure Complexity} & \textbf{Accuracy \& Generality} & \textbf{Mathematical Tractability} \\ \midrule
			\raisebox{-0.5\height}{\begin{tabular}{c}Ray-tracing \\ models \cite{metis,3gpp.38.901,itu-r} \end{tabular}}  & \textbullet\ High computational cost \newline \textbullet\ Unable for link-level {\color{white} \textbullet\textbullet}simulations & \textbullet\ High accuracy of capturing ELAA spatial {\color{white} \textbullet\textbullet}non-stationarities \newline \textbullet\ Environment and layout dependent & \textbullet\ Weak tractability and {\color{white} \textbullet\textbullet}hardly support analytical {\color{white} \textbullet\textbullet}study\\
			\midrule
			\raisebox{-0.5\height}{\begin{tabular}{c}GBSMs \cite{metis, 3gpp.38.901, itu-r, Flordelis2020,9120570,9257469,9318511,Bian2018,Jaeckel2019,Rodrigues2020,Yang2019}\\ (e.g, COST 2100) \end{tabular}} & \textbullet\ Moderate computational cost \newline \textbullet\ Unsuitable for link-level {\color{white} \textbullet\textbullet}simulations & \textbullet\ Moderate accuracy of capturing key ELAA {\color{white} \textbullet\textbullet}spatial non-stationarities \newline \textbullet\ Statistical environment knowledge dependent & \textbullet\ Moderate tractability \newline \textbullet\ Highly complex for {\color{white} \textbullet\textbullet}analytical study \\
			\midrule
			\raisebox{-0.4\height}{\begin{tabular}{c} Existing ELAA spatially \\ non-stationary fading \\ channel models \cite{Amiri2018,1510955,Wang2022,8638522,Amiri2022,8949454}  \end{tabular}} & \textbullet\ Low computational cost \newline \textbullet\ Suitable for link-level {\color{white} \textbullet\textbullet}simulations & \textbullet\ Low accuracy and fail to capture some of key {\color{white} \textbullet\textbullet}ELAA spatial non-stationarities \newline \textbullet\ Over-simplified and loss of generality & \textbullet\ Strong tractability and {\color{white} \textbullet\textbullet}suitable for analytical {\color{white} \textbullet\textbullet}study \\
			\bottomrule[1.5pt]
		\end{tabular}}
\vspace{-0em}	
\end{table*}
Current massive-MIMO channel models mainly include ray-tracing models, \glspl{gbsm}, and fading channel models (i.e., correlation-based stochastic models) \cite{Wang2018}.
For the sake of saving space, we only provide a brief review of massive-MIMO channel models in \tabref{tabR1}, which shows that fading channel models are more suitable for link-level simulations and analytical studies due to their simpler structures \cite{BJORNSON20193}.

In the literature, the simplest spatially non-stationary fading channel models are independent and non-identically distributed (i.n.d.) Rice (or Rayleigh) (e.g., \cite{Amiri2018,1510955,Wang2022}).
Their p.d.f. varies with respect to the index pair $(m,n)$, i.e.,
\begin{subequations}\label{eqn05}
	\begin{align}
		&f_{H_{m,n}}(x)=\mathcal{CN}(x; \mu_{m,n}, \sigma_{m,n}^2);\label{eqn5a}\\
		&\quad\mu_{m,n}^*\mu_{m,n}+\sigma_{m,n}^2=\frac{\rho}{d_{m,n}^{\alpha}}\label{eqn5b},
	\end{align}
\end{subequations}
where the distance $d$ varies with the index pair $(m,n)$. 
Specially for the i.n.d. Rayleigh channel, Eqn. \eqref{eqn5b} reduces to
\begin{equation}\label{eqn06}
	\mu_{m,n}=0,~\sigma_{m,n}^2=\frac{\rho}{d_{m,n}^{\alpha}},~\forall m,n.
\end{equation}

In the literature \cite{8638522,Amiri2022,8949454}, a generalized version of i.n.d. Rayleigh channel is considered by incorporating the concept of visibility region. 
Mathematically, the idea is to add a binary random variable ($\eta_{m,n}\in\{0,1\}$) on top of \eqref{eqn5a}, i.e.,
\begin{equation}\label{eqn07}
	H_{m,n}\left\{
	\begin{array}{l}
		\sim\mathcal{CN}(x; \mu_{m,n}, \sigma_{m,n}^2);~\eta_{m,n}=1\\
		=0; ~\eta_{m,n}=0
	\end{array}
	\right.
\end{equation}
where $\eta_{m,n}$ is not independently distributed with respect to the index pair $(m,n)$.
To facilitate our presentation in the rest of the paper, we denote $[\cdot]_m$ to be the antenna index on the ELAA side and $[\cdot]_n$ the antenna index on the UT side.
Then, the $n^{th}$ column of $\mathbf{H}$ (denoted by $\mathbf{h}_n\in\mathbb{C}^M$) is the channel vector between the ELAA and the $n^{th}$ antenna on the UT side. 
Correspondingly, we have a column vector $\mat{\eta}_n=[\eta_{0,n}, ...,\eta_{M-1,n}]^T$, with $[\cdot]^T$ denoting the vector/matrix transpose.
Usually, $\mat{\eta}_n$ consists of a number of $\mathbf{1}$-vectors and $\mathbf{0}$-vectors.
Those $\mathbf{1}$-vectors are corresponding to the visibility regions.
It is often assumed that $\mathbf{1}$-vector has its length obeying log-normal distribution and centre obeying uniform distribution.
In \cite{Flordelis2020,9257469,9318511,9120570}, those $\mathbf{1}$-vectors are generated using the birth-death process or Markov process.

In our preliminary work \cite{Liu2021}, the concept of visibility region is faded out. 
Instead, each element of $\mathbf{H}$ is described as the relationship of three random variables ($\epsilon_{m,n},  \widetilde{H}_{m,n}, \beta_{m,n}$), i.e.,
\begin{equation}\label{eqn08}
	H_{m,n}=\epsilon_{m,n}^{(\beta_{m,n})} \widetilde{H}_{m,n}^{(\beta_{m,n})},
\end{equation}
where $\beta_{m,n}\in\{0,1\}$ is a binary random variable, with $\beta_{m,n}=1$ indicating the link on the LoS state or otherwise $\beta_{m,n}= 0$ indicating the NLoS state.
One major contribution of this paper is the random distribution of $\beta$, which will be discussed in detail throughout \secref{sec2};
$\epsilon_{m,n}^{(\beta_{m,n})}$ represents the log-normal distributed shadowing effect, where $\epsilon_{m,n}^{(1)} \sim \mathcal{LN}(0, \sigma_{\epsilon,\textsc{l}}^2)$ and $\epsilon_{m,n}^{(0)} \sim \mathcal{LN}(0, \sigma_{\epsilon,\textsc{n}}^2)$. Here, $\sigma_{\epsilon,\textsc{l}}^2$ and $\sigma_{\epsilon,\textsc{n}}^2$ denote the variance of $\ln(\epsilon)$ in LoS and NLoS state, respectively \cite{3gpp.38.901};
$\widetilde{H}_{m,n}^{(\beta_{m,n})}\sim f_{H_{m,n}}(x)$.
More specifically, $\widetilde{H}_{m,n}^{(0)}$ obeys i.n.d. Rayleigh distribution as follows \cite{Liu2023, Liu2023b}
	\begin{equation} \label{eqnR9}
		\widetilde{H}_{m,n}^{(0)} \triangleq \bigg(\sqrt{\dfrac{\rho_{\textsc{n}}}{d_{m,n}^{\alpha_{_\textsc{n}}}}} \ \bigg) \delta_{m,n},
	\end{equation}  
where $\delta_{m,n} \sim \mathcal{CN}(0,1)$ denotes the complex Gaussian random variable;
$\rho_{\textsc{n}}$ and $\alpha_{\textsc{n}}$ denote the path-loss coefficient and exponent of NLoS links, respectively.
$\widetilde{H}_{m,n}^{(1)}$ obeys i.n.d. Rice distribution as follows \cite{Liu2023a}
\begin{equation} \label{eqnR10}
	\widetilde{H}_{m,n}^{(1)} \triangleq \bigg(\sqrt{\dfrac{\rho_{\textsc{l}}}{d_{m,n}^{\alpha_{_\textsc{l}}}}} \ \bigg) \bigg(\sqrt{\dfrac{\kappa}{\kappa + 1}}\phi_{m,n} + \sqrt{\dfrac{1}{\kappa + 1}}\delta_{m,n}\bigg),
\end{equation} 
where $\phi_{m,n} = \exp(-j\frac{2\pi}{\lambda}d_{m,n})$ denotes the phase of the LoS path, $\lambda$ the carrier wavelength, $\kappa \sim \mathcal{LN}(\mu_{\kappa},\sigma^{2}_{\kappa})$ is the Rice K-factor \cite{3gpp.38.901}, and $\rho_{\textsc{l}}$ and $\alpha_{\textsc{l}}$ stand for the path-loss coefficient and exponent of LoS links, respectively.
The spherical-wave model indicates that the RSS varies from link to link as a function of the distance \cite{7414041}.
This characteristic is properly captured by the first terms in equations \eqref{eqnR9} and \eqref{eqnR10}.
This model is the generalized version of \eqref{eqn07} by means of: 
{\it 1)} replacing the binary random variable $\eta_{m,n}$ with the continuous random variable $\epsilon_{m,n}$. 
The issue of visible region is captured by evaluating whether the RSS of user-to-service antenna links exceeds a certain threshold (see \cite[Fig. 2]{Liu2021});
{\it 2)} introducing the binary random variable $\beta_{m,n}$ to enable a mix of LoS/NLoS states in the fading channel model.
The random distribution of $\beta_{m,n}$ has been preliminarily studied in \cite{Liu2021}, which was mainly focused on the single-user equivalent scenario.

\subsection{Contribution of This Work}
In this paper, we aim to provide a comprehensive study on the fading channel model specified in \eqref{eqn08}. 
In addition to the single-user equivalent model presented in \cite{Liu2021}, this work is mainly focused on the probability distribution of $\beta_{m,n}$ in the multiuser fading channel model. 
Our idea is to divide \glspl{ut} into two groups: a group of reference users (RUs) and a group of non-RUs.
The definition of RUs and non-RUs is stated as:
\begin{defn}\label{def01}
	Consider each UT having a single antenna. 
	Form a column vector $\mat{\beta}_n=[\beta_{0,n},...,\beta_{M-1,n}]^T$, which indicates the LoS/NLoS condition of links between the $n^{th}$ UT and the ELAA. 
	Denote $\mathbf{\Phi}$ to be the set of RUs and $\overline{\mathbf{\Phi}}$ the set of non-RUs with $\mathbf{\Phi}\cap\overline{\mathbf{\Phi}}=\emptyset$.
	UTs which belong to the set $\overline{\mathbf{\Phi}}$ are called non-RUs. 
	For any two UT indices $n_1, n_2\in\mathbf{\Phi}$, the joint probability of $\mat{\beta}_{n_1}, \mat{\beta}_{n_2}$ fulfills:
	\begin{equation}\label{eqn09}
		p(\mat{\beta}_{n_1}, \mat{\beta}_{n_2})=p(\mat{\beta}_{n_1})p(\mat{\beta}_{n_2}).
	\end{equation}
\end{defn}

For RUs (i.e., the condition \eqref{eqn09} holds), $\mat{\beta}_n$ can be independently generated $\forall n\in\mathbf{\Phi}$. 
The algorithm used to generate $\mat{\beta}_n$ can be found in \cite{Liu2021} as well as in Section \ref{sec2a} of this paper.
For non-RUs, $\mat{\beta}_k, \forall k\in\overline{\mathbf{\Phi}}$, is correlated with some of $\mat{\beta}_n, \forall n\in\mathbf{\Phi}$.
Then, a stochastic orthonormal-basis-expansion (S-OBE) method is proposed to generate $\mat{\beta}_k, \forall k\in\overline{\mathbf{\Phi}}$ (see Section \ref{sec2b}). 
It is demonstrated that, with appropriate parameter configurations, the proposed spatially non-stationary fading channel model can yield computer-simulated channel data that matches the published data measured from practical ELAA-MIMO channels (see \cite{Harris2016} and \cite{7063445}). 

With these appealing results, the proposed spatially non-stationary fading channel model is employed to study the cumulative distribution function (CDF) of ELAA-MIMO channel capacity. 
The CDF is obtained through parametric fitting and linear regression in various typical scenarios (see \secref{sec3}). 
For all of our studied scenarios, it is unveiled that the ELAA-MIMO channel capacity obeys the skew normal distribution.
Moreover, the channel capacity is also found close to the Gaussian or Weibull distribution, depending on the geo-location and distribution of UTs.
More specifically, for the single-user equivalent scenario or multiuser scenarios with short UT-to-ELAA distances (e.g., $1\ \mathrm{m}$), the channel capacity is close to the Gaussian distribution; and for others, it is close to the Weibull distribution.
Compared to the skew normal distribution, the latter two distributions have the advantage of simpler structures, which facilitate the (semi-) analytical study.
These are important results which can help wireless engineers to quickly predict the outage probability of ELAA-MIMO channels and find out optimized solutions towards practical wireless problems.

Finally, with the proposed spatially non-stationary fading channel model, we investigate the impact of channel spatial non-stationarity on linear MIMO receivers mainly through computer simulations. 
It is demonstrated that linear MIMO receivers can offer near-optimum detection performance in WSSUS massive-MIMO channels can be sub-optimum (e.g., about $2$ dB) in spatially non-stationary channels.
This calls for low-complexity and near-optimum ELAA transceivers for future MIMO technology.

The rest of of this paper is organized as follows. 
Section \ref{sec2} presents the proposed spatially non-stationary fading channel model. 
Section \ref{sec3} presents the (semi-) analytical study of the ELAA-MIMO channel capacity.
Section \ref{sec4} evaluates linear MIMO receivers in the proposed spatially non-stationary fading channel.
Finally, the conclusion and outlook are presented in Section \ref{sec5}.

\section{Novel spatially Non-stationary Fading Channel Model for ELAA-MIMO}\label{sec2}
The focus of this section is on the random distribution of $\beta_{m,n}, \forall m,n,$ in the spatially non-stationary channel model \eqref{eqn08}.
Other involved random variables ($\epsilon_{m,n}, \widetilde{H}_{m,n}$) have their random distribution already determined and they are independently generated for different UTs. 
It is worth noting that antennas co-located on the same UT share the same LoS/NLoS condition.
For the sake of concise presentation, we stress that UTs only have a single antenna. 
Then, the term $\mat{\beta}_n$ simply denotes LoS/NLoS condition for the $n^{th}$ UT.
In the proposed channel model, $\mat{\beta}_n$ for RUs is independently generated  by employing exponentially decaying windows \cite{Liu2021}.
Then, $\mat{\beta}_n$ for non-RUs is randomly generated by utilizing the S-OBE method.
With this in mind, we will discuss in detail how to randomly generate $\mat{\beta}_n$ for RUs and non-RUs, respectively. 

\subsection{Generating LoS/NLoS State for RUs ($\mat{\beta}_n\in\mathbf{\Phi}$)}\label{sec2a}
{\it Definition \ref{def01}} implies that RUs (i.e., $n\in\mathbf{\Phi}$) can have their $\mat{\beta}_n$ independently generated with respect to $n$. 
This immediately turns a multiuser problem into a single-user equivalent problem. 
For the sake of notation simplicity, we drop the subscript $[\cdot]_n$ in $\mat{\beta}_n$ for the rest of discussion in this subsection.
The stochastic behavior of $\mat{\beta}$ is summarized in the following result (from \cite[\em Proposition 1]{Liu2021}):
\begin{thm}\label{thm01}
	Suppose that $\beta_\ell, \ell\in\{0,...,M-1\},$ obeys the Bernoulli distribution:
	\begin{equation}\label{eqn10}
		\mathcal{B}(\beta_\ell; \varrho_\ell)=\varrho_\ell^{\beta_{\ell}}(1-\varrho_\ell)^{1-\beta_\ell},~\beta_\ell\in\{0,1\},
	\end{equation}
	where $\varrho_\ell=p(\beta_\ell=1)$.
	The probability mass function (PMF) of $\beta_m,~_{\forall m\neq\ell},$ is $\mathcal{B}(\beta_m;\varrho_m)$ with $\varrho_m$ given by
	\begin{equation}\label{eqn11}
		\varrho_m=(2\varrho_\ell-1)\exp\Big(-\frac{\Delta_{\ell,m}}{d_{\textsc{los}}}\Big)+1-\varrho_\ell,
	\end{equation}
	where $\Delta_{\ell,m}$ is the distance between two service-antennas indexed by $\ell, m$, respectively, 
	and $d_\textsc{los}$ is the correlation distance of LoS state which is defined in the \gls{3gpp} document \cite{3gpp.38.901}.
\end{thm}
\begin{IEEEproof}
	{\em Theorem \ref{thm01}} gets well supported by the measurement results in \gls{3gpp} and \gls{itu-r} technical documents \cite{3gpp.38.901,itu-r}. 
	In \cite[(Table 7.4.2-1)]{3gpp.38.901}, the probability of LoS state is described as a function of the \gls{2d} distance ($d^\mathrm{2D}_m$) between service-antennas and UTs. 
	For instance, in the propagation environment of \gls{umi}, the probability for a link to be LoS is specified by
	\begin{IEEEeqnarray}{ll}\label{eqn12}
		p_{\textsc{los}}(d^\textsc{2d}_m) = & \min\left(\frac{\overline{d_1}}{d^\textsc{2d}_m},1\right) \left[1-\exp\left(-\frac{d^\textsc{2d}_m}{\overline{d_2}}\right)\right] \nonumber\\
		&\quad\quad\quad\quad\quad\quad\quad\quad+ \exp \left(-\frac{d^\textsc{2d}_m}{\overline{d_2}}\right),
	\end{IEEEeqnarray}
	where $\overline{d_1}$, $\overline{d_2}$ are the reference distances that can be found in \cite{3gpp.38.901}, as well as $p_{\textsc{los}}(d^\textsc{2d}_m)$ for various environments.
	It is worth noting that when the $d_{m}^{\textsc{2d}}$ falls below $\overline{d_1}$, the LoS probability becomes $1$.
	The feasibility of using 3-D distances in \eqref{eqn12} has been suggested as a potentially more practical option, but there is presently a shortage of accessible data to support this approach.
	Regardless, as this is not the main contribution of this paper, we can temporarily use 2-D distance as a provisional solution.
	For the $\ell^{th}$ service antenna, its LoS probability in \eqref{eqn11} is given by
	\begin{equation}\label{eqn13}
		\varrho_\ell=p_{\textsc{los}}(d^\textsc{2d}_\ell).
	\end{equation}
	Moreover, the probability for two service-antennas to share the same LoS/NLoS state is also specified in \cite[(7.4-5) and Table 7.6.3.1-2]{3gpp.38.901}, i.e.,
	\begin{equation}\label{eqn14}
		p_\textsc{los}(\ell, m) = \exp\left(-\frac{\Delta_{\ell, m}}{d_\textsc{los}}\right).
	\end{equation}
	When the ELAA is a \gls{ula}, \eqref{eqn14} becomes
	\begin{equation}\label{eqn15}
		p_\textsc{los}(\ell, m) = \exp\left(-\frac{\lambda|\ell-m|}{2d_\textsc{los}}\right),
	\end{equation}
	With \eqref{eqn12} and \eqref{eqn14}, we can compute the LoS probability $(p_m)$ for the $m^{th}$ antenna as
	\begin{IEEEeqnarray}{ll}
		\varrho_m & =p_\textsc{los}(\ell, m)\varrho_\ell+(1-\varrho_\ell)(1-p_\textsc{los}(\ell, m)), \nonumber \\
		& = 2 p_\textsc{los}(\ell, m)\varrho_\ell -p_\textsc{los}(\ell, m) + 1 -\varrho_\ell , \nonumber  \\
		& =(2 \varrho_\ell -1)p_\textsc{los}(\ell, m) + 1 -\varrho_\ell. \label{eqn16}
	\end{IEEEeqnarray}
	Plugging \eqref{eqn14} into \eqref{eqn16}, we can obtain the result \eqref{eqn11}.
\end{IEEEproof}

\begin{cor}[Corollary 1 in \cite{Liu2021}]\label{cor1}
	Suppose that the binary random variable $\beta_m$ obeys the correlated Bernoulli distribution specified in [{\em Theorem \ref{thm01}}, \eqref{eqn11}].
	For the \gls{ula}, the number of adjacent service-antennas sharing the same LoS (or NLoS) state (denoted by $L$) obeys the following random distribution
	\begin{equation}\label{eqn17}
		f_L(x)= \exp\left(\dfrac{\lambda(x^2 -x)}{-4d_{\textsc{los}}}\right) - \exp\left(\dfrac{\lambda(x^2 + x)}{-4d_{\textsc{los}}}\right).
	\end{equation}
\end{cor}
\begin{IEEEproof}
	See \appref{appA}
\end{IEEEproof}

{\it Remark 1.1:}
{\em Theorem \ref{thm01}} shows that the correlation in \gls{los}/NLoS state between two service-antennas decreases exponentially with their distance 
$\Delta_{\ell, m}$.
This effectively forms an exponentially decaying window for antenna correlation, and the length of which (i.e., $L$) follows the distribution specified in {\it Corollary \ref{cor1}}.

{\it Remark 1.2:} This channel model is also suitable for the ELAA-MIMO systems deployed with a rectangular antenna-array (URA). 
As discussed in \cite{BJORNSON20193}, a LoS/NLoS window can be generated from the lowest row of the URA. 
For those antennas located in the same column, if the lowest placed antenna is on the LoS state, then the others in the column are also on the LoS state. 
Or otherwise, if a lower antenna is on the NLoS state, then the antenna located one row above can be generated according to the Bernoulli distribution \eqref{eqn06}. 

{\it Remark 1.3:}
According to 3GPP \cite{3gpp.38.901}, \eqref{eqn10} can also be employed to describe the spatial inconsistency of shadowing effects, where the term $d_\textsc{los}$ is replaced by $d_\textsc{sf}$ (the correlation distance of shadowing factor).
Usually, $d_\textsc{sf}$ is smaller than $d_\textsc{los}$. 
Moreover, it is often assumed that, within the same shadowing window, service-antennas share the same shadowing effect.

Finally, as a summary of the above theoretical discussion, $\mat{\beta}$ for RUs can be generated using the pseudocode provided in {\bf Algorithm \ref{alg01}} (see Appendix \ref{appB}).

\subsection{Generating LoS/NLoS State for non-RUs ($\mat{\beta}_k\in\overline{\mathbf{\Phi}}$)}\label{sec2b}
Here, our objective is to generate $\mat{\beta}_k, \forall k\in\overline{\mathbf{\Phi}}$.
This is generally a challenging task since $\mat{\beta}_k$ can be statistically correlated with each other and they are also correlated with $\mat{\beta}_n, \forall n\in\mathbf{\Phi}$. 
For the sake of modeling simplicity, our fading channel model is based on the following assumptions: 

{\it A1):} $\mat{\beta}_k, \forall k\in \overline{\mathbf{\Phi}},$ are conditionally independent, i.e.,
\begin{equation}\label{eqn18}
	p(\mat{\beta}_k, \forall k|\mat{\beta}_n, \forall n)=\prod_{k\in\overline{\mathbf{\Phi}}}p(\mat{\beta}_k|\mat{\beta}_n, \forall n),
\end{equation}
with which $\mat{\beta}_k$ is only dependent on $\mat{\beta}_n, \forall n\in\mathbf{\Phi}$.
Then, our study can be mainly focused on $p(\mat{\beta}_k|\mat{\beta}_n, \forall n)$.

{\it A2):} Consider a pair $(\mat{\beta}_k, \mat{\beta}_n), k\in\overline{\mathbf{\Phi}}, n\in\mathbf{\Phi}$. 
They have identical element-wise cross-correlation, i.e.,
\begin{equation}\label{eqn19}
	\mathcal{C}_{\beta}(k,n;m)=\mathrm{cor}(\beta_{m,k},\beta_{m,n})=\zeta_{k,n}~(\mathrm{const.}), \forall m.
\end{equation}
Moreover, they share the same auto-correlation, i.e.,
\begin{IEEEeqnarray}{ll}\label{eqn20}
	\mathcal{C}_{\beta}(k;m,\tau)&=\mathrm{cor}(\beta_{m,k},\beta_{m+\tau,k})\\
	&=\mathrm{cor}(\beta_{m,n},\beta_{m+\tau,n})=\mathcal{C}_{\beta}(n; m,\tau),\label{eqn21}
\end{IEEEeqnarray}
where $\tau$ is the lag of auto-correlation.
Our experimental results in {\it Case Study 2} confirm that {\it A1} and {\it A2} are valid assumptions for practical ELAA-MIMO scenarios.

\begin{thm}\label{thm02}
	To fulfill the assumptions {\it A1)} and {\it A2)}, the following linear model of $\mat{\beta}_k$ forms a sufficient condition
	\begin{equation}\label{eqn22}
		\mat{\beta}_k=\zeta_{k,n}\mat{\beta}_n+\mat{\omega}_n, \forall n\in\mathbf{\Phi},
	\end{equation}
	where $\mat{\omega}_n$ is an independent random process of $\mat{\beta}_n$ with 
	\begin{equation}\label{eqn23}
		\mathrm{cor}(\omega_{m_1,n_1},\omega_{m_2,n_2})=0,~{m_1\neq m_1, n_1\neq n_2}.
	\end{equation}
\end{thm}
\begin{IEEEproof}
	Firstly, \eqref{eqn22} shows that $\mat{\beta}_k$ is only dependent on $\mat{\beta}_n$ but not any other non-RUs. 
	Hence, the assumption {\it A1)} is straightforwardly satisfied. 
	Next, we perform element-wise cross-correlation between $\mat{\beta}_k$ and $\mat{\beta}_n$ which yields
	\begin{IEEEeqnarray}{ll}
		\mathcal{C}_{\beta}(k,n;m)&=\mathrm{cor}(\beta_{m,k},\beta_{m,n})\nonumber\\
		&=\mathrm{cor}(\zeta_{k,n}\beta_{m,n}+\omega_{m,n},\beta_{m,n})\nonumber\\
		&=\underbrace{\mathrm{cor}(\zeta_{k,n}\beta_{m,n},\beta_{m,n})}_{=\zeta_{k,n}}+\underbrace{\mathrm{cor}(\omega_{m,n},\beta_{m,n})}_{=0}; \ 
	\end{IEEEeqnarray}
	\begin{IEEEeqnarray}{ll}\label{eqn25}
		&\mathcal{C}_{\beta}(k;m,\tau)=\mathrm{cor}(\zeta_{k,n}\beta_{m,n},\zeta_{k,n}\beta_{m+\tau,n})\nonumber\\
		&\quad\quad+\mathrm{cor}(\zeta_{k,n}\beta_{m,n},\omega_{m+\tau,n})+\mathrm{cor}(\zeta_{k,n}\beta_{m+\tau,n},\omega_{m,n})\nonumber\\
		&\quad\quad+\mathrm{cor}(\omega_{m,n},\omega_{m+\tau,n}).
	\end{IEEEeqnarray}
	Given the property of $\mat{\omega}_n$ specified in {\it Theorem \ref{thm02}}, the last three terms at the right hand of Eqn. \eqref{eqn25} vanish. 
	Then, \eqref{eqn25} is equivalent to \eqref{eqn21}.
	{\em Theorem \ref{thm02}} is therefore proved. 
\end{IEEEproof}

With {\em Theorem \ref{thm02}}, we can represent $\mat{\beta}_k$ in the following form
\begin{equation}\label{eqn26}
	\mat{\beta}_k=\dfrac{1}{J}\left(\sum_{n\in\mathbf{\Phi}}\zeta_{k,n}\mat{\beta}_n+\sum_{n\in\mathbf{\Phi}}\mat{\omega}_n\right),
\end{equation}
where $J$ denotes the cardinality of $\mathbf{\Phi}$.
{W}e form a $(J)\times(1)$ vector $\mat{\zeta}_k=[\zeta_{k,0},\dots ,\zeta_{k,J-1}]^T$
and an $(M)\times(J)$ matrix $\mathbf{B}$ by collecting $\mat{\beta}_n, \forall n\in\mathbf{\Phi}$.
Then, the following result is reached:
\begin{cor}\label{cor21}
	A sufficient condition to fulfill the assumptions {\em A1)} and {\em A2)} is to represent $\mat{\beta}_k$ with the following S-OBE form
	\begin{equation}\label{eqn27}
		\mat{\beta}_k=\frac{1}{J}\mathbf{B}\mat{\zeta}_k+\mat{\omega}_k,
	\end{equation}
	where $\mat{\omega}_k=\frac{1}{J}\sum_{n\in\mathbf{\Phi}}\mat{\omega}_n$.
	More concisely, the term $(1)/(J)$ can be incorporated into $\mat{\zeta}_k$ and \eqref{eqn27} has a more compact form
	\begin{equation}\label{eqn28}
		\mat{\beta}_k=\mathbf{B}\mat{\zeta}_k+\mat{\omega}_k.
	\end{equation}
\end{cor}

In \eqref{eqn28}, $\mathbf{B}$ describes the LoS/NLoS states of RUs which can be obtained through {\bf Algorithm \ref{alg01}}.
The correlation vector $\mat{\zeta}_k$ is dependent on specific radio environment. 
Nonetheless, $\mat{\zeta}_k$ fulfills two conditions: 
{\em c1)} $0\leq\zeta_{k,n}\leq 1$; {\em c2)} $\sum_{n=0}^{J-1}\zeta_{k,n}=1,$
with which every element of $[\mathbf{B}\mat{\zeta}_k]$ falls into the range of $[0, 1]$. 
{\it Corollary \ref{cor21}} does not specify the random distribution of $\mat{\omega}_k$ and this gives us sufficient flexibility to define the stochastic characteristic of $\mat{\beta}_k$.

\begin{cor}\label{cor22}
	Define $\mat{\varrho}_k=\mathbf{B}\mat{\zeta}_k$.
	There exists a random vector $\mat{\omega}_k$ such that: {\em 1)} $\mathbb{E}(\mat{\beta}_k)=\mat{\varrho}_k$; 
	{\em 2)} the $m^{th}$ element of $\mat{\beta}_k$ has the variance of $\varrho_{m,k}(1-\varrho_{m,k})$, 
	where $\mathbb{E}(\cdot)$ stands for the expectation and $\varrho_{m,k}$ for the $m^{th}$ element of $\mat{\varrho}_k$.
\end{cor}
\begin{IEEEproof}
	A sufficient condition for $\mathbb{E}(\mat{\beta}_k)=\mat{\varrho}_k$ is: $\mathbb{E}(\mat{\omega}_k)=\mathbf{0}$.
	Under this condition, $\beta_{m,k}$ and $\omega_{m,k}$ share the same variance. 
	If we set $\mathbb{E}(\omega_{m,k}^*\omega_{m,k})=\varrho_{m,k}(1-\varrho_{m,k})$, {\it Corollary \ref{cor22}} is proved. 
\end{IEEEproof}

Concerning $\beta_{m,k}\in\{0,1\}$, a suitable random distribution that fulfills {\em Corollary \ref{cor22}} is: 
\begin{equation}\label{eqn29}
	\beta_{m,k}\sim\mathcal{B}(\beta_{m,k};\varrho_{m,k}).
\end{equation}
In computer simulations, the major obstacle to implement \eqref{eqn29} lies in: 
{\em 1)} to determine the set of RUs ($\mathbf{\Phi}$); 
{\em 2)} to determine the correlation vector $\mat{\zeta}_k, \forall k\in\overline{\mathbf{\Phi}}$. 
Both are dependent on the network architecture, the size of network as well as the radio environment. 
In this paper, we provide a case study of ELAA-MIMO based on the linear Wyner-type model, i.e., the Wyner cellular channel model illustrated in \cite[Fig.1]{5990396}, where the base-stations are replaced with the ELAA.
\begin{figure}[t]
	\centering
	\includegraphics[width=0.95\linewidth]{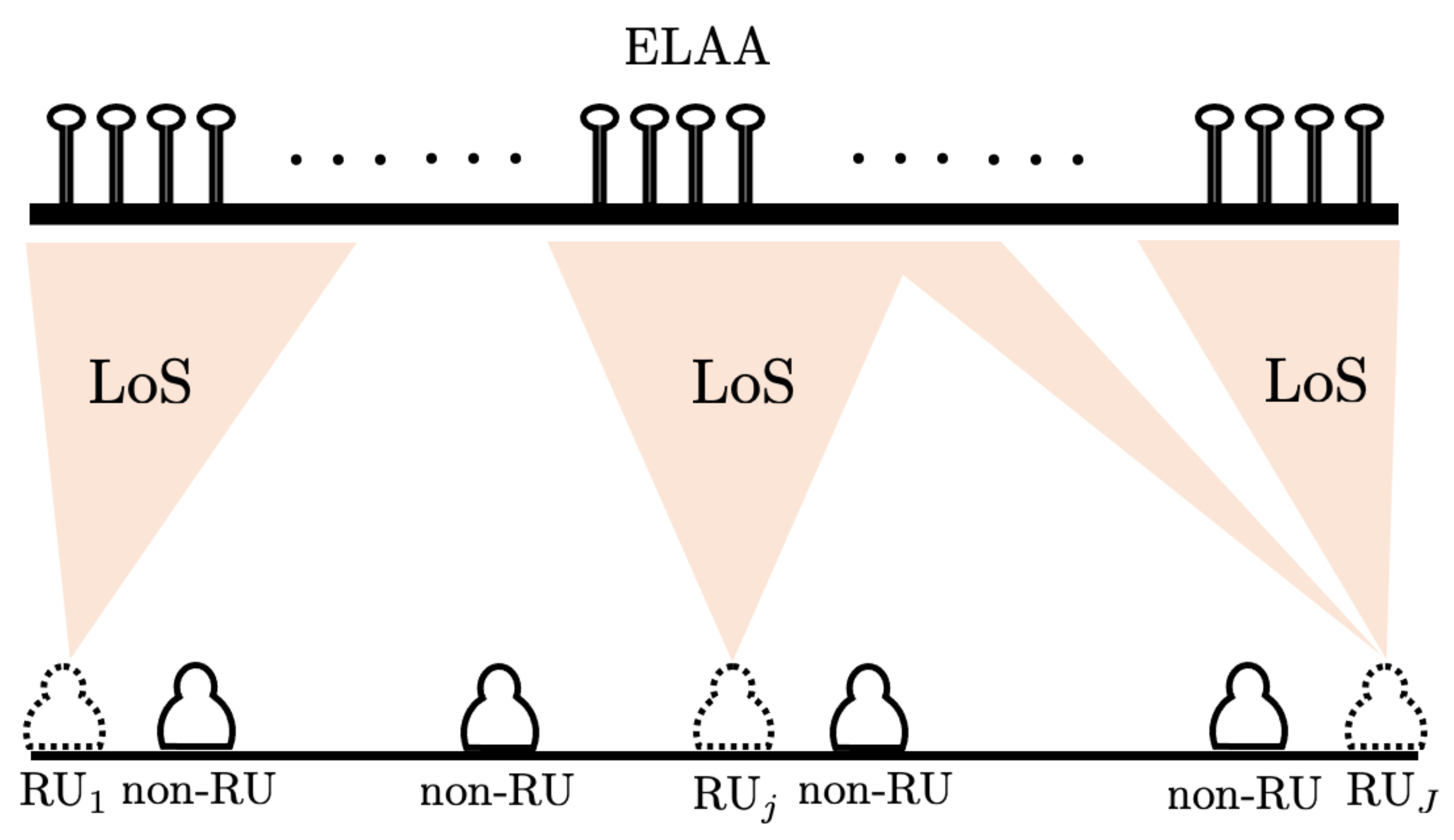}	
	\vspace{-1em}
	\caption{An example of linear Wyner-type model of ELAA-MIMO system. In the proposed channel model, NLoS links can also experience the issue of visible region when the RSS of a link exceeds a certain threshold.}
	\label{fig01}
	\vspace{-0em}
\end{figure}

As illustrated in Fig. \ref{fig01},  there are a number of RU places ($J$) on the line in parallel to the ELAA.
Note that RU places are different from RUs in the sense that RUs do not need to be present on those places.
Those RU places are indexed by the subscript $[\cdot]_j$ in the ascent order (i.e., $j=1,...,J$) all the way from the left-hand side to the right. 
Assume a UT placed between RU$_j$ and RU$_{j+1}$. 
The LoS/NLoS state of this UT ($\mat{\beta}_k$) will be only correlated with RU$_j$ ($\mat{\beta}_j$) and RU$_{j+1}$ ($\mat{\beta}_{j+1}$);
or otherwise, $\mat{\beta}_j$ and $\mat{\beta}_{j+1}$ cannot be independently generated. 
Hence, for the linear Wyner-type model, $\mat{\varrho}_k$ reduces to
\begin{equation}\label{eqn30}
	\mat{\varrho}_k=\sum_{n=j}^{j+1}\zeta_{k,n}\mat{\beta}_n.
\end{equation}
For the correlation coefficients $\zeta_{k,j}$ and $\zeta_{k,j+1}$, we propose the following configuration
\begin{equation}\label{eqn31}
	\zeta_{k,j}=\frac{d_{j,k}}{d_{j,j+1}}, ~\zeta_{k,j+1}=\frac{d_{j+1,k}}{d_{j,j+1}},
\end{equation}
where $d_{j,k}$ denotes the distance between RU$_{j}$ and the UT;
$d_{j,j+1}$ denotes the distance between RU$_{j}$ and RU$_{j+1}$ ($d_{j,j+1}=d_{j,k}+d_{j+1,k}$).
For the special case of $d_{j,k}=0$ or $d_{j+1,k}=0$, the UT is actually a RU; 
and for other cases, the UT is a non-RU.

As a summary of the above discussion, $\mat{\beta}_k$ for non-RUs can be generated using the pseudocode provided in {\bf Algorithm \ref{alg02}} (see Appendix \ref{appC}). 
As long as $\mat{\beta}_n$ is obtained for all UTs, we can employ \eqref{eqn08} to randomly generate the channel matrix $\mathbf{H}$.

\subsection{Channel Normalization}
In computer simulations, wireless MIMO channels must be appropriately normalized for the sake of fair comparison. 
When the MIMO channel is WSSUS, the channel matrix is often normalized by $\mathbb{E}\|\mathbf{H}\|^2$, 
which can be computed by
\begin{equation}\label{eqn32}
	\mathbb{E}\|\mathbf{H}\|^2 = \sum_{m=0}^{M-1}\sum_{n=0}^{N-1}\mathbb{E}\vert H_{m,n}\vert^2=\mathrm{const.},
\end{equation}
where $\|\cdot\|$ stands for the Frobenius norm, and $|\cdot|$ stands for the modulus.
However, due to the spatial non-stationarity of ELAA-MIMO channels, $H_{m,n}$ can have different p.d.f. for different Monte Carlo trial. 
This slightly complicates the channel normalization procedure. 

With \eqref{eqn08}, we compute $\mathbb{E}|H_{m,n}|^2$ as 
\begin{IEEEeqnarray}{ll}
	\mathbb{E}|H_{m,n}|^2&=\varrho_{m,n}\mathbb{E}|\epsilon_{m,n}^{(1)}|^2\mathbb{E}|H_{m,n}^{(1)}|^2\nonumber\\
	&\quad\quad+(1-\varrho_{m,n})\mathbb{E}|\epsilon_{m,n}^{(0)}|^2\mathbb{E}|H_{m,n}^{(0)}|^2.
\end{IEEEeqnarray}
The probability ($\varrho_{m,n}$) varies with each Monte Carlo trial. 
Therefore, we suggest the use of \eqref{eqn13} as an approximation.  
Moreover, given the shadow effect following the log-normal distribution, we have \cite{Holgate1989}
\begin{equation}\label{eqn34}
	\mathbb{E}|\epsilon_{m,n}^{(1)}|^2=\exp\Big(\frac{\sigma_{\epsilon,\textsc{l}}^2}{2}\Big),~ \mathbb{E}|\epsilon_{m,n}^{(0)}|^2=\exp\Big(\frac{\sigma_{\epsilon,\textsc{n}}^2}{2}\Big),
\end{equation}
Finally, the terms ($\mathbb{E}|H_{m,n}^{(1)}|^2,\mathbb{E}|H_{m,n}^{(0)}|^2$) can be computed by employing \eqref{eqn5b}.

\subsection{Showcases of the Proposed spatially Non-stationary Fading Channel Model}\label{sec2d}
\subsubsection*{Case Study 1}
\begin{figure*}[t]
	\centering
	\subfigure[LoS/NLoS state]{	
		\centering	
		\includegraphics[height=4.5cm]{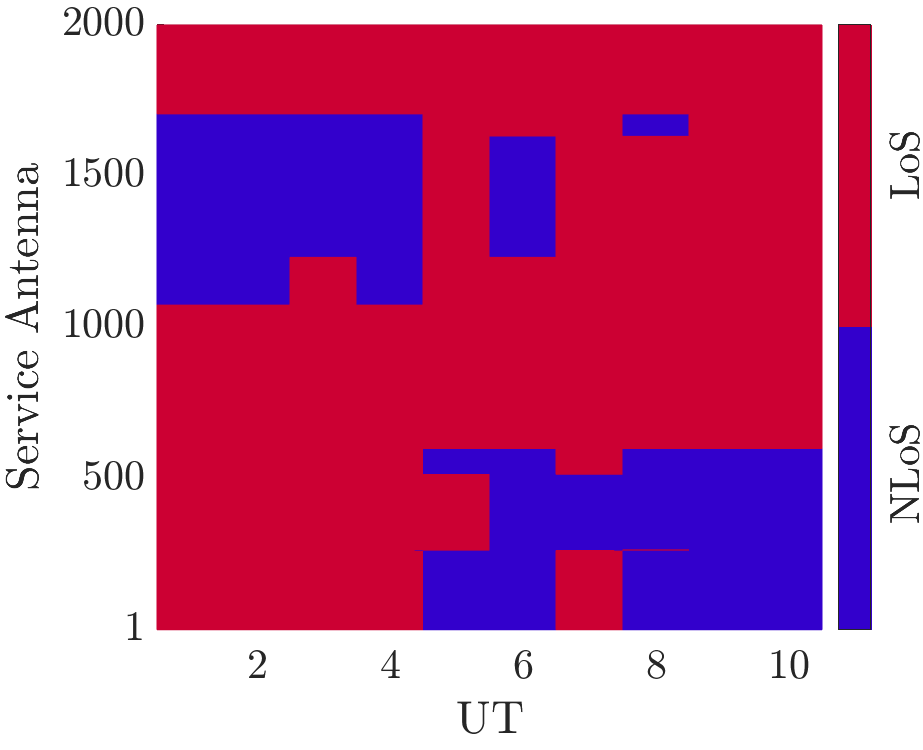}
		\label{fig02a}
	}
	\hfil
	\subfigure[RSS\quad\quad\quad\quad]{
		\centering	
		\includegraphics[height=4.5cm]{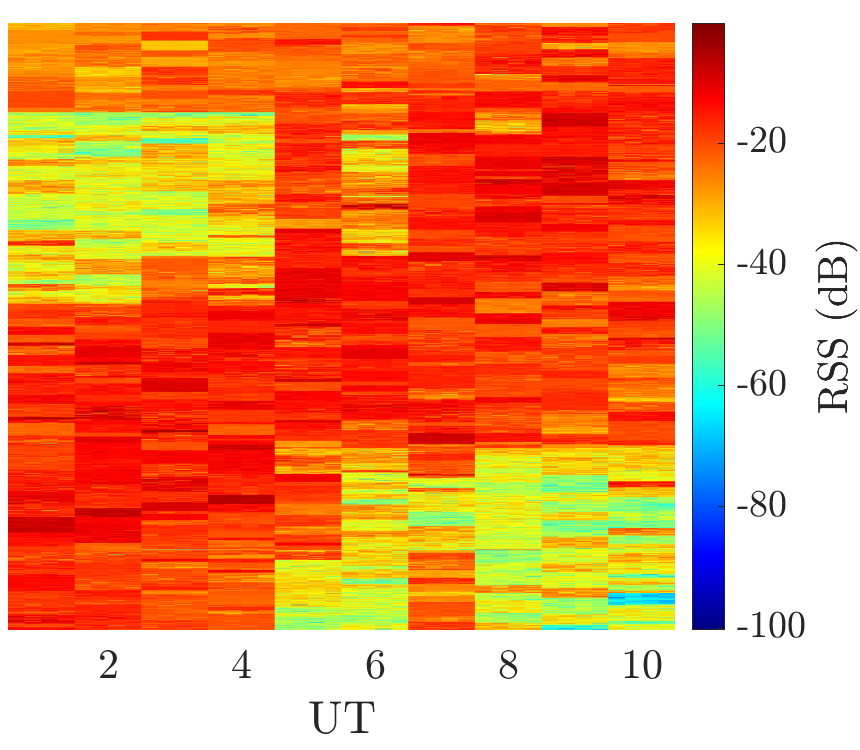}
		\label{fig02b}
	}
	\vspace{-0.5em}
	\caption{\label{fig02} LoS/NLoS state and RSS from UTs to ELAA antennas. UT $\texttt{\#}1$ and UT $\texttt{\#}10$ are RUs and UT $\texttt{\#} 2$ - $\texttt{\#} 9$ are non-RUs. (a): the LoS/NLoS state of every link. (b): the RSS (normalized by the transmitted power) heat map.}
	\vspace{-0em}
\end{figure*}
This case study aims to demonstrate the spatial correlation and non-stationarity of the proposed fading channel model through computer simulations.
UTs and the ELAA are deployed according to the linear Wyner-type model as shown in \figref{fig01}.
In the simulation, the ELAA is a large ULA, which has $M = 2,000$ service-antennas and the length of $85$ m (assuming $3.5$ GHz central frequency).
There are $10$ UTs (including $2$ RUs and $8$ non-RUs), with each having $N_{\textsc{ue}} = 4$ antennas, placed in parallel to the ELAA with equal spacing.
The two RUs are placed at the two ends of the UT side.
The perpendicular distance between UTs and ELAA, denoted by $d_{\perp}$, is set as $25$ m.
The propagation environment is the \gls{umi}-street canyon.
According to the \gls{3gpp} document \cite{3gpp.38.901}, the system parameters are configured by: $d_{\textsc{los}} = 5000$, $d_{\textsc{sf}} = 15$, $\overline{d_1} = 18$, $\overline{d_2} = 36$, $\rho_{\textsc{l}} = 0.007$, $\rho_{\textsc{n}} = 0.020$, $\alpha_{\textsc{l}} = 1.050$,  $\alpha_{\textsc{n}} = 1.765$, $\sigma_{\epsilon,\textsc{l}} = 4$ $\mathrm{dB}$, $\sigma_{\epsilon,\textsc{n}} = 7.82$ $\mathrm{dB}$, $\mu_\kappa = 9$ $\mathrm{dB}$, and $\sigma_{\kappa} = 10$ $\mathrm{dB}$.

For a single random channel realization, the LoS/NLoS state and RSS of every UT-to-ELAA antenna link are illustrated in \figref{fig02}.
In practical environments, the LoS/NLoS states of RU places are typically considered independent because they are located at a sufficient distance from each other.
To simplify the case study, it is assumed that UT $\texttt{\#} 1$ and UT $\texttt{\#} 10$ are the two RUs, as they are the farthest apart among all the users.
\figref{fig02a} shows that non-RUs have their LoS/NLoS states more similar to their nearest RU.
For instance, UT $\texttt{\#} 2$ (non-RU) shares the same LoS/NLoS state as UT $\texttt{\#}1$ (RU).
It is also demonstrated that, for some places where the two RUs have the same LoS/NLoS state (e.g., $m \in [600, 1074]$ or $m \in [1706, 2000]$), all the corresponding links of non-RUs will share the identical LoS/NLoS state.
\figref{fig02b} shows the RSS (normalized by the transmitted power) of every UT-to-ELAA antennas link.
Due to $d_{\textsc{sf}} < d_{\textsc{los}}$, it can be observed that shadowing effects vary more rapidly than LoS/NLoS states.
Moreover, LoS links demonstrate higher RSS than NLoS links. 
This phenomenon coincides with practical wireless propagation channels.

\subsubsection*{Case Study 2}
In this case study, the objective is to illustrate that the proposed channel model can be utilized to generate computer-simulated channel data that closely align with two distinct sets of measurement data obtained from practical ELAA-MIMO channels.
One measurement data-set is obtained from the published materials \cite{Harris2016},
where their channel measurement campaign was carried out in an indoor environment with mostly LoS conditions. 
More specifically (please see the floor plan in \cite{Harris2016}), there is an ULA having $M = 112$ antennas with the aperture of $5.4$ $\mathrm{m}$.
Moreover, there are $12$ single-antenna UTs placed on a line in parallel to the ULA. 
UTs are placed with the inter-UT-distance of $50$ $\mathrm{cm}$. 
The perpendicular distance between UTs and ELAA is: $d_\perp=12$ $\mathrm{m}$.
For the above parameters, we use the same setting in our computer simulations.  
For other parameters which are not specified in \cite{Harris2016}, we use the recommendation in the 3GPP document for indoors \cite{3gpp.38.901}:  $d_{\textsc{sf}} = 5$, $\rho_{\textsc{l}} = 0.007$, $\alpha_{\textsc{l}} = 0.865$, $\sigma_{\epsilon,\textsc{l}} = 3$ $\mathrm{dB}$, $\mu_{\kappa} = 7$ $\mathrm{dB}$, and $\sigma_{\kappa} = 8$ $\mathrm{dB}$.

\begin{figure*}[t]
	\centering
	\subfigure[Measurement result in \cite{Harris2016} \quad \quad \quad (b) The proposed model \quad \quad \quad \quad \quad \quad \quad \quad (c) Absolute distance \quad \quad]{	
		\centering	
		\includegraphics[height=4cm]{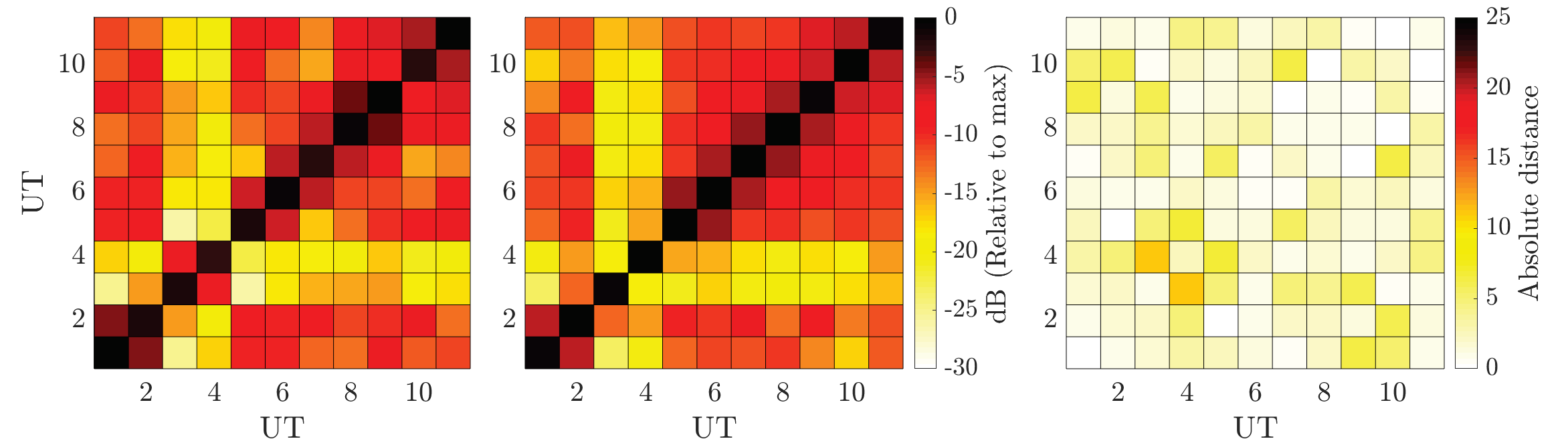}}
	\vspace{-0em}
	\caption{Comparison of $\mathbf{H}\mathbf{H}^H$ intensity (normalized by the maximum element) between (a) the measured results in \cite{Harris2016} and (b) the proposed ELAA channel model. The absolute distance between (a) and (b) is shown in (c).}
	\label{fig03}
	\vspace{-0em}
\end{figure*}
\begin{figure*}[t]
	\centering
	\subfigure[Measurement result in \cite{7063445} \quad \quad \quad (b) The proposed model \quad \quad \quad \quad \quad \quad \quad \quad (c) Absolute distance \quad \quad]{	
		\centering	
		\includegraphics[height=4cm]{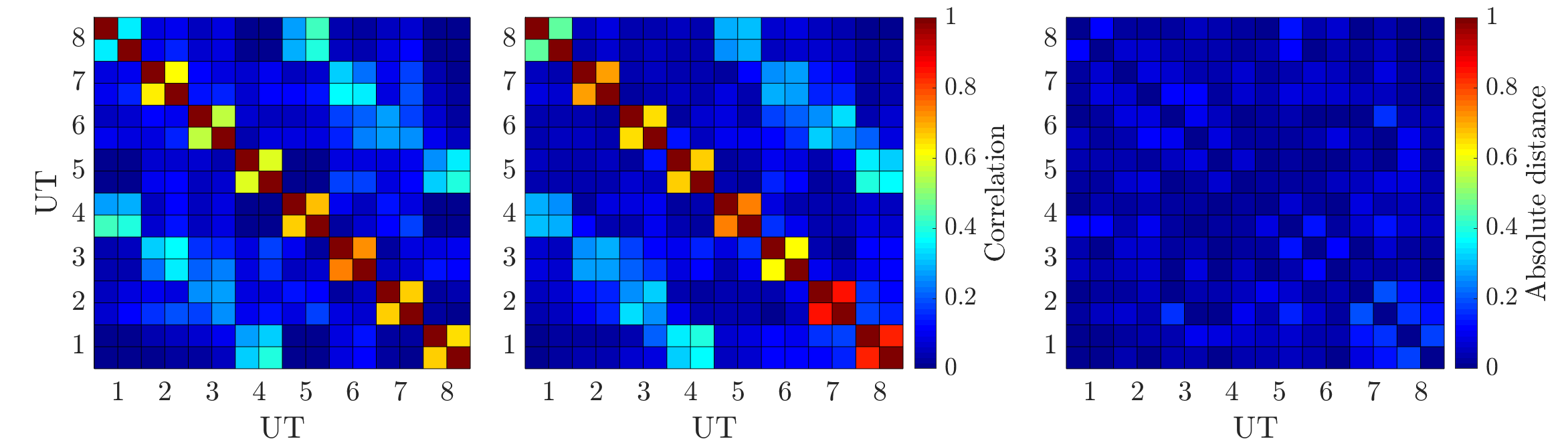}}
	\caption{Comparison of channel correlations between (a) the measured results in \cite{7063445} and (b) the proposed ELAA channel model. The absolute distance between (a) and (b) is shown in (c).}
	\label{fig03R}
	\vspace{-0em}
\end{figure*}
\figref{fig03} shows the $\mathbf{H}\mathbf{H}^{H}$ intensity of the measured (see \cite[Fig. 12]{Harris2016}) and computer-simulated channel data, where all illustrated values are normalized by the maximum intensity.
\figref{fig03}(c) shows the absolute distance between \figref{fig03}(a) and \figref{fig03}(b).
First of all, it can be observed that the proposed fading channel model can yield computer-simulated channel data that matches well with the measurement data.
This is evident in \figref{fig03}(c), where the absolute distance between measured and computer generated data is relatively small.
There is a notable but minor difference observed between two pictures in the figure, i.e., those channel data related to UT $\texttt{\#}3$ and UT $\texttt{\#} 4$.
This is because the two UTs have their small-scale fading correlated in the real-channel measurement.
However,  the correlation of small-scale fading is not incorporated in our current channel model. 
We recognize the significance of considering small-scale correlation in MIMO channel models, but would argue that the incorporation of this correlation is straightforward.
To focus our discussion on the key novelty of this work, the correlation of small-scale fading is not emphasized in this paper. 


The other set of measurement data is acquired from the published resources found in \cite{7063445}.
The referenced study employs a ULA with $M = 64$ antennas and a large aperture of $2$ $\mathrm{m}$. 
It also involves $8$ UTs, each containing $2$ antennas, positioned parallel to the ULA across two rows (see the floor plan in \cite{7063445}).
The same settings are adopted for the aforementioned parameters in our computer simulations. The measurement environment in that study is \gls{umi}-street canyon. Therefore, for parameters not specified in \cite{7063445}, the settings correspond to those in \textit{Case Study 1}.
The same setting is used for the above parameters in our computer simulations. 
The measurement environment in that study is \gls{umi}-street canyon. 
Therefore, for the parameters not specified in \cite{7063445}, the same settings as in {\it Case Study 1} are used.
\figref{fig03R}(a) and \figref{fig03R}(b) depict the UT-antennas correlations for both the measured (see \cite[Fig. 6(b)]{7063445}) and computer-simulated channel data. As shown in \figref{fig03R}(c), the absolute distance between these correlations is small, indicating that the proposed channel model can effectively represent the spatially non-stationary characteristics of practical ELAA channels.

\section{CDF of ELAA-MIMO Channel Capacity with Massive Service-Antennas}\label{sec3}
\subsection{MIMO Channel Capacity and Statistical Behavior}
Assuming the channel-state information ($\mathbf{H}$) to be available only at the receiver side, the conditional channel capacity (or sometimes called instantaneous channel capacity) under transmit-power constraint is given by \cite{1175470,4460100604,1203954}
\begin{equation}\label{eqn35}
	C|_\mathbf{H}=\log _{2}\det \left(\mathbf{I}_{M}+\frac{\gamma_o}{N} \mathbf{H} \mathbf{H}^{H}\right) \quad \mathrm{bit} / \mathrm{s} / \mathrm{Hz},
\end{equation}
where $\gamma_o$ is the average \gls{snr} at each individual receive-antenna.

For fading channels, the probability of capacity outage is given by
\begin{equation}\label{eqn36}
	P_\mathrm{out}=\int_0^{C_\mathrm{out}}f_{C}(x)dx,
\end{equation}
where $f_{C}(x)$ is the p.d.f of channel capacity, and $C_\mathrm{out}$ is the outage capacity.
When the massive-MIMO channel is WSSUS, the capacity exhibits channel hardening effect (see \cite{Hochwald2004})
\begin{equation}\label{eqn37}
	\lim_{M\rightarrow\infty}\frac{C|_\mathbf{H}}{\mathbb{E}(C)}=1,
\end{equation}
i.e., the conditional channel capacity is asymptotically equal to the unconditional channel capacity. 
When the outage capacity is set to $C_\mathrm{out}=\mathbb{E}(C)$, we have $P_\mathrm{out}=0$ for any rate that is lower than $\mathbb{E}(C)$. However, our preliminary result presented in \cite{Liu2021} shows that \eqref{eqn37} does not necessarily hold when the massive-MIMO fading channel is spatially non-stationary. 
The following section is therefore motivated. 

\subsection{CDF of The ELAA-MIMO Channel Capacity with The Proposed Model}
The CDF of ELAA-MIMO channel capacity is defined as follows
\begin{equation}
	F_{C}(x) = p(C < x),
\end{equation}
which can be used to predict the outage probability. 
However, due to the spatial non-stationarity, it is difficult to analytically describe $F_{C}(x)$ for the ELAA-MIMO channel.
Therefore, in the rest of this paper, our analysis is focused on the empirical CDF (ECDF) of the channel capacity.
Denote $T$ as the total number of Monte Carlo trials, the ECDF is defined as follows
\begin{equation}
	\widehat{F}_{T}(x) = \frac{1}{T}\sum_{t = 1}^{T}\mat{i}\{C_{t} \leq x\},
\end{equation}
where $C_t$ represents the channel capacity of the $t^{th}$ Monte Carlo trial, and $\mat{i}\{\cdot\}$ is the indicator of the input event.
According to the Glivenko–Cantelli theorem \cite{Tucker1959}, $\widehat{F}_{T}(x)$ converges to $F_{C}(x)$, with $100\%$ probability, when $T$ tends to be infinity, i.e.,
\begin{equation}\label{eqn40}
	\sup _{T \rightarrow \infty} \big|\widehat{F}_{T}(x) - F_{C}(x)\big| \rightarrow 0 \text{ almost surely},
\end{equation}
where $\sup|\cdot|$ denotes the supremum norm. 
This means that the analysis of $F_{C}(x)$ is equivalent to the analysis of $\widehat{F}_{T}(x)$ when $T$ is sufficiently large.

Usually, mean (denoted by $\mu_{T}$) and variance (denoted by $\sigma^2_{T}$) are required to characterize the distribution of channel capacity. 
In practice, these two parameters are replaced by their ensemble averages, respectively:
\begin{subequations}
	\begin{equation}
		\mu_{T}  = \dfrac{1}{T}\left(\sum_{t = 1}^{T} C_t\right);
	\end{equation}
	\begin{equation}
		\sigma_{T}^{2} = \frac{1}{T}\left(\sum_{t = 1}^{T}\left(C_{t} - \mu_{T}\right)^{2}\right),
	\end{equation}
\end{subequations}
where $\sigma_{T}^2$ reflects the fluctuation of channel capacity.
For WSSUS MIMO channels, \eqref{eqn37} implies that $\sigma^2_{T}$ tends to be $0$ when $M$ tends to be infinity.
However, our numerical results (in Sec. \ref{sec3e}) will show that this is not the case for ELAA-MIMO channels. 
In short, $\sigma^2_{T}$, in ELAA-MIMO channels, exhibits
\begin{eqnarray}
	\lim_{M \rightarrow \infty}\sigma_{T}^{2} \rightarrow k_o ~(\mathrm{const.}),
\end{eqnarray}
where $k_o$ is considerably large (see \figref{fig04}).
It again confirms the significance of studying the stochastic behavior of channel capacity even when the ELAA-MIMO system is equipped with very massive service-antennas (e.g., $M=2,000$ or higher). 
Parametric fitting and linear regression are the prominent methods employed in our study.

\subsection{Parametric Fitting of The ELAA-MIMO Channel Capacity}
In this section, the objective is to find some statistical models which are suitable to describe the distribution of ELAA-MIMO channel capacity \cite{Freedman2009}.
The first step of parametric fitting is to assume that the channel capacity obeys a statistical distribution with a p.d.f. denoted by $f_{X}(x; \mat{\theta})$, where $\mat{\theta}$ is the parameter set.
The number of parameters dependents on the structure of the assumed distribution.
Then, the parameter set is determined by minimizing the following negative log-likelihood function \cite{Rossi2018}
\begin{equation}\label{eqn44}
	\widehat{\mat{\theta}} = \argmin_{\mat{\theta}} -\sum_{t = 1}^{T} \ln f_{X}\left(C_t; \mat{\theta}\right),
\end{equation}
which is so-called \gls{mle}, and $\widehat{\mat{\theta}}$ is the estimated parameter set.
However, the negative log-likelihood value is not suitable for measuring the fitting error of different statistical models, as it does not satisfy the triangle inequality \cite{Geisser2006,Canonne2022}.
Moreover, the CDF of ELAA-MIMO channel capacity is more significant for predicting the outage probability.
Therefore, the following residual is proposed to measure the error of parametric fitting
\begin{equation}
	\Theta = \sqrt{\sum_{t = 1}^{T} \left(\widehat{F}_{T}(C_t) - F_{X}(C_t| \widehat{\mat{\theta}})\right)^2},
\end{equation}
where $\Theta$ denotes the error of parametric fitting, and $F_{X}(x| \widehat{\mat{\theta}})$ is the CDF of the assumed distribution with the estimated parameter set.

In this paper, the ECDF of ELAA-MIMO channel capacity is fitted by multiple statistical models (see \appref{appR1}) in various typical ELAA scenarios (see \secref{sec3e}).
In every scenario, the skew normal distribution is found to have the lowest fitting error with the following CDF \cite{Mudholkar2000}
\begin{equation}
	F^{\textsc{sn}}_{X}(x; \mat{\theta}) =  \Psi\left(\dfrac{x-\theta_1}{\theta_2}\right)  - 2 \mathcal{T}\left(\dfrac{x-\theta_1}{\theta_2}, \theta_3\right),
\end{equation}
where $\Psi(\cdot)$ is the p.d.f. of standard Gaussian distribution, which is defined as follows
\begin{equation}
	\Psi(x) = \dfrac{1}{\sqrt{2\pi}}\exp\left(-\frac{x^2}{2}\right),
\end{equation}
and $\mathcal{T}(\cdot)$ is Owen's T function as follows \cite{Owen1956}
\begin{equation}
	\mathcal{T}\left(\theta_1, \theta_2\right) = \frac{1}{2 \pi} \int_{0}^{\theta_2} \frac{\exp \left({-\frac{1}{2}\theta_{1}^{2}\left(1+x^{2}\right)}\right)}{1+x^{2}} d x,
\end{equation}
where $-\infty < \theta_1, \theta_2 < +\infty$ are the two parameters of the function.

Moreover, it is also found that, for single-user equivalent scenarios or multiuser scenarios with short $d_{\perp}$, the channel capacity is close to the Gaussian distribution whose CDF follows \cite{Squires2001}
\begin{equation}\label{eqn49}
	F^{\textsc{gs}}_{X}(x; \mat{\theta}) = \dfrac{1}{2} \left[1 + \operatorname{erf} \left(\dfrac{x-\theta_1}{\sqrt{2}\theta_2} \right) \right],
\end{equation}
where $\operatorname{erf}(\cdot)$ is the error function \cite{Andrews1998}; $\theta_1$ and $\theta_2$ are the mean and standard deviation, respectively.
When $T$ tends to be infinity, the estimated parameters of the Gaussian distribution are as follows \cite{Anderson1985}
\begin{equation}
	\widehat{\theta}_1 = \mu_{T}, \text{ and } \lim_{T\rightarrow\infty}\widehat{\theta}_2 = \sigma_{T}.
\end{equation}
For multiuser scenarios with relative large $d_\perp$, the ELAA-MIMO channel capacity is found close to the Weibull distribution whose CDF is as follows \cite{Papoulis2001}
\begin{equation}
	F_{X}^{\textsc{wb}}(x; \mat{\theta}) = 1 -  \exp\left(-\left({x} / {\theta_1}\right)^{\theta_2}\right),
\end{equation}
where $\theta_1$ and $\theta_2$ are the scale and shape parameters, respectively.

Compared to the latter two distributions, the skew normal distribution has the advantage of higher accuracy.
The latter two distributions have some loss of accuracy and generality compared to the skew normal distribution, but they have the advantage of simpler structures, which facilitates the (semi-) analytical study.
Since $\gamma_o$ is a continuous variable, it is impractical to perform parametric fitting at all possible average-SNRs.
This motives the following subsection on linear regression.

\subsection{Linear Regression of  Every Element in $\widehat{\mat{\theta}}$}
To simplify the notation, $\widehat{\theta}$ is denoted to be any of the estimated parameter (e.g., $\widehat{\theta}_{1}, \widehat{\theta}_2$ of the Gaussian distribution).
After performing parametric fitting at $R$ ($R \geq 2$) different $\gamma_o$, the relationship between $\gamma_o$ and $\widehat{\theta}$ is found to be approximately linear as follows
\begin{equation}\label{eqn52}
	\widehat{\theta} \approx  a\gamma_o + c,
\end{equation}
where $a$, $c$ are two coefficients.
The values of these two coefficients can be determined by linear regression, whose objective function is defined as follows \cite{Montgomery2021}
\begin{equation}\label{eqn53}
	(a, c) = \underset{(a, c) \in \mathbb{R}}{\arg \min }\sum_{r = 1}^{R} \left(a \gamma_r + c - \widehat{\theta}_{\gamma_r}\right)^{2}.
\end{equation}
where $\gamma_r$ and $\widehat{\theta}_{\gamma_r}$ denote the $r^{th}$ average-SNR and the corresponding estimated parameter, respectively.

\subsection{Numerical Results and Discussion}\label{sec3e}
In this section, the objectives are: {\em 1)} to study the channel hardening effect of ELAA-MIMO channel capacity; {\em 2)} to showcase the results of parametric fitting, and {\em 3)} to showcase the results of linear regression.
The propagation environment is configured with the same parameters as in {\em Case Study 1}.
For every experiment, the number of Monte Carlo trials is set as $T = 10,000$.
The following three cases of the proposed fading channel model are considered:

{\bf Case 1} ($1$ RU and $4$ non-RUs): all the UTs have the same LoS/NLoS state, which usually occurs when UTs are close to each other in practice (high density). Therefore, all the UTs are deployed within a $2$-meter range in this case;

{\bf Case 2} ($2$ RUs and $3$ non-RUs): the two RUs are placed at the two ends of the UT side.
This is a medium density case where the UTs are deployed within a 20-meter range;

{\bf Case 3} ($5$ RUs): this is the single-user equivalent scenario where the LoS/NLoS states of UTs are generated independently.
This is a low density case where the UTs are deployed within a $50$-meter range.

From {\bf Case 1} to {\bf Case 3}, the correlation between UTs becomes weaker.
The objectives of this section set the following three case studies for our numerical study.

\begin{figure}[t]
	\centering
	\includegraphics[width=0.45\textwidth]{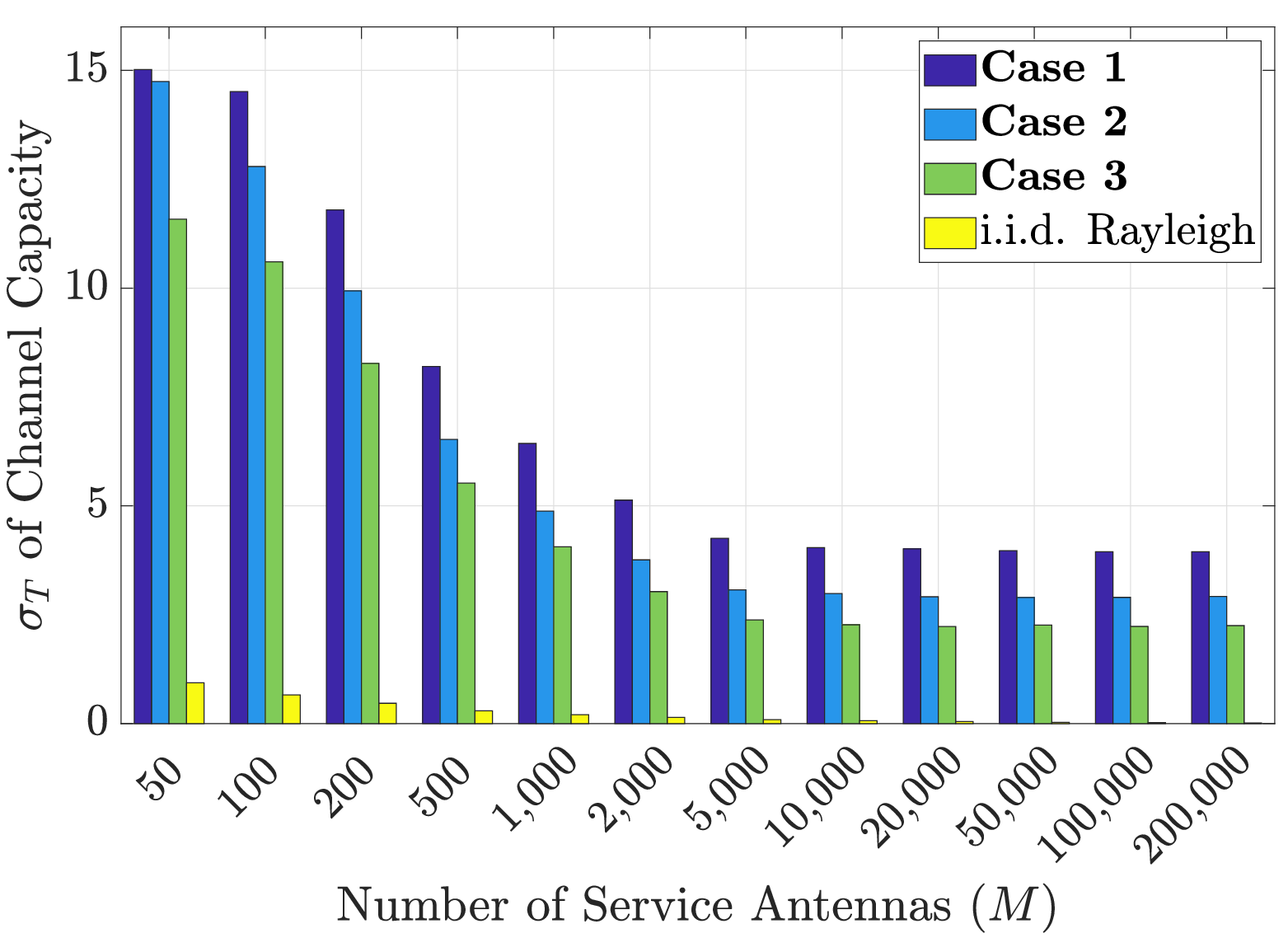}
	\vspace{-0.5em}
	\caption{Standard deviation ($\sigma_{T}$) of the channel capacity. $\gamma_o = 10$ $\mathrm{dB}$; $N = 20$; $d_{\perp} = 50$ m; and $T=10,000$.}
	\label{fig04}
	\vspace{-1em}
\end{figure}
{\em Case Study 3:}
The objective of this case study is to demonstrate the fluctuation of ELAA channel capacity, ranging from practical to extreme cases.
\figref{fig04} shows $\sigma_{T}$ of the channel capacity instead of $\sigma_{T}^2$, because $\sigma^{2}_{T}$ is too small to be observed in the i.i.d. Rayleigh channels.
In this figure, there are $N = 20$ ($N_{\textsc{ue}} = 4$) UT antennas and the ratio of $M/N$ is set from $10$ to $10,000$, which means that the maximum ELAA aperture is about $8$ $\mathrm{km}$.
As shown in the figure, when $M$ is relatively small, $\sigma_{T}$ becomes smaller as $M$ increases.
It can be observed that the channel hardening effects exists in conventional MIMO systems with i.i.d. Rayleigh channels, i.e., $\sigma_T \approx 0$ when $M$ is significantly large.
However, for ELAA-MIMO channels, $\sigma_{T}$ converges to a constant when $M$ becomes sufficiently large (e.g., $M \geq 10,000$).
This means that the channel hardening effect does not hold in the ELAA-MIMO system.
Moreover, with the same $M$, $\sigma_{T}$ becomes smaller from {\bf Case 1} to {\bf Case 3}.
This means that the correlation of LoS/NLoS states between different
UTs can render the ELAA-MIMO channel more non-stationary.
To explain the nonexistence of the channel hardening effect in ELAA-MIMO systems, we have designed the following case study.

\begin{figure*}[t]
	\centering
	\subfigure[\quad\quad \quad \quad \quad \quad \quad \quad \quad \quad \quad \quad \quad \quad \quad \quad \quad \quad \quad \quad \quad (b) \quad \quad \quad \quad \quad \quad \quad \quad \quad \quad \quad \quad \quad \quad \quad \quad \quad (c)]{	
		\centering	
		\includegraphics[width=0.93\textwidth]{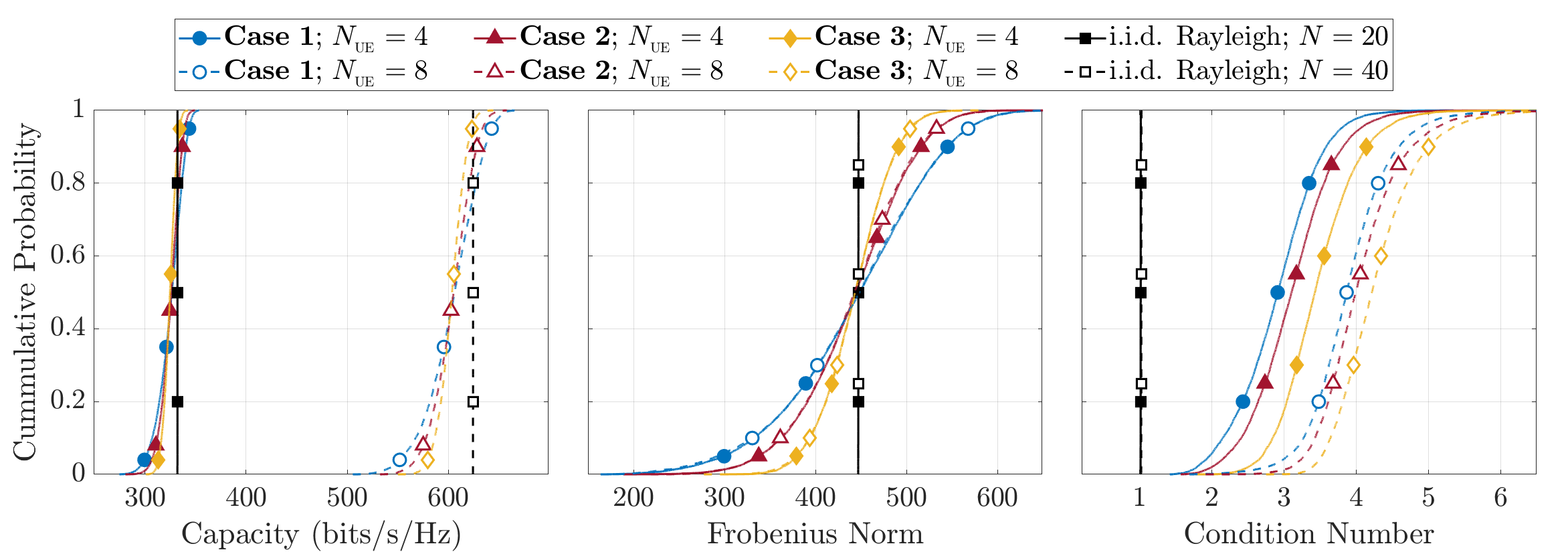}}
	\caption{ECDF of (a) the channel capacity, (b) Frobeniu s norm, and (c) condition number at average-SNR $=10$ $\mathrm{dB}$, $d_{\perp} = 50$ m, $T=10,000$, and $M = 200,000$.}
	\label{fig05}
	\vspace{-0em}
\end{figure*}
\begin{figure*}[t]
	\centering
	\subfigure[$d_\perp = 1$ $\mathrm{m}$]{	
		\centering	
		\includegraphics[width=0.45\textwidth]{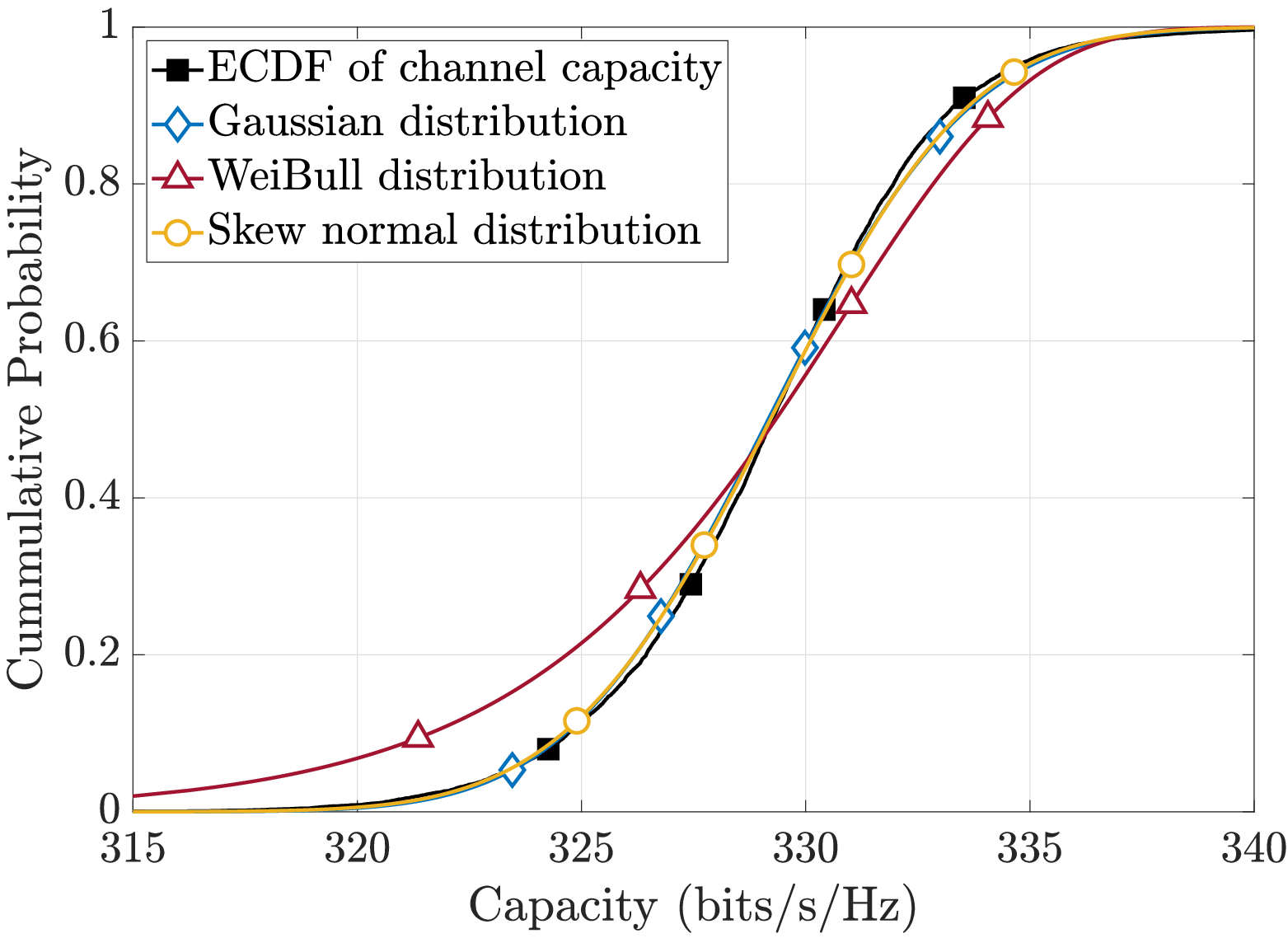}
		\label{fig06a}
	}
	\subfigure[$d_\perp = 50$ $\mathrm{m}$]{
		\centering	
		\includegraphics[width=0.45\textwidth]{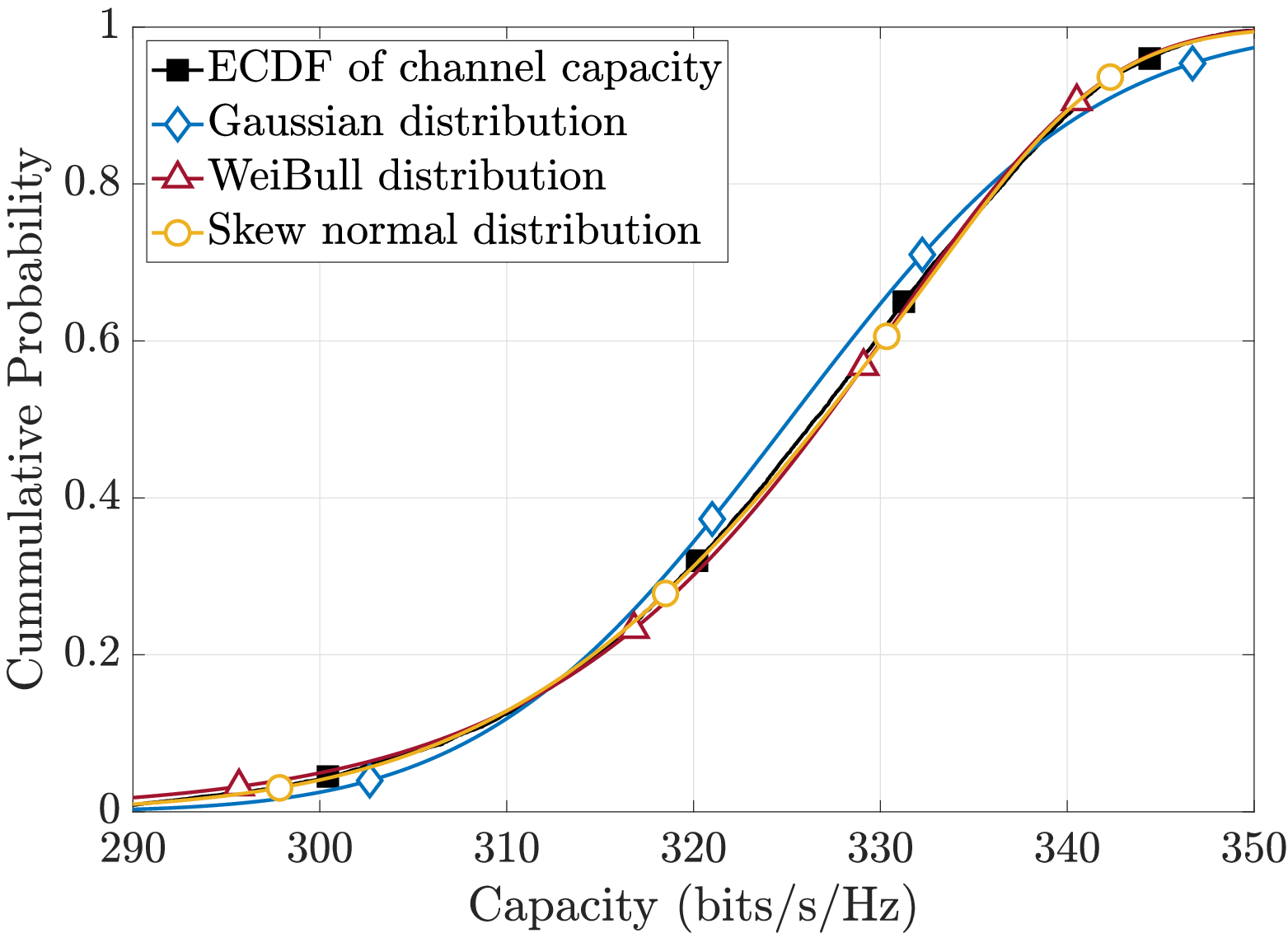}
		\label{fig06b}
	}
	\caption{\label{fig06}Parametric fitting results of the Gaussian, Weibull and skew normal distributions in two scenarios with different UT-to-ELAA distances ($d_\perp$). {\bf Case 1}; $\gamma_o = 10$ $\mathrm{dB}$; $N_{\textsc{ue} }= 4$; $M = 200,000$.}
	\vspace{-0.5em}
\end{figure*}
{\em Case Study 4:}
The objective of this case study is to demonstrate the reasons behind the disappearance of channel hardening effects in ELAA-MIMO systems.
In stationary MIMO channels, the channel hardening effects is observable when $M$ is sufficiently large, as described in \eqref{eqn37}.
Therefore, in this case study, we consider an extremely large $M$ to enrich our understanding of the theoretical distribution of ELAA channel capacity.
The ECDF of channel capacity is also plotted in \figref{fig05}(a), where average-SNR ($\gamma_o$) is set as $10$ $\mathrm{dB}$ for all the exhibited channels.
In MIMO systems, the primary determinants of the channel capacity are the channel norm and channel condition.
Their statistical behaviors in various MIMO channels are shown in \figref{fig05}(b) and \figref{fig05}(c).
In i.i.d. Rayleigh fading channels, the channel hardening effects can be observed, primarily due to the channel norm and condition number remain almost constant across different channel realizations.
This finding aligns with many references highlighting the channel hardening effects in stationary MIMO channels, e.g, \cite{Hochwald2004,fowc,Du2010,2009wireless,BJORNSON20193}.

Contrary to conventional MIMO systems, the channel hardening effect is not applicable in the context of ELAA-MIMO systems, as illustrated in Fig. 5(a).
The spatial non-stationarity intrinsic to the ELAA channel significantly influences the stability of the channel norm and condition number, thereby leading to obvious fluctuations. 
Such fluctuations disrupt the typical channel hardening effect observed in stationary MIMO systems.
This has been explored in sources such as \cite{7063445} and \cite{Wang2018}, as well as in our prior publications \cite{Liu2021}.
Moreover, to study the impact of intra-UT correlation, the results of $N_\textsc{ue} = 4, 8$ are shown in this figure.
\figref{fig05}(b) shows the ECDF of the channel Frobenius norm.
The fluctuations of the norm become smaller from {\bf Case 1} to {\bf Case 3}, which implies that the inter-UT correlation weakens the channel hardening effects.
\figref{fig05}(c) shows the ECDF of the channel condition number.
It shows that both the inter-UT and intra-UT correlations can render the ELAA-MIMO channel more ill-conditioned.
It can be found that the statistical behavior of the channel capacity well coincides with the statistical behaviors of the channel Frobenius norm and condition number.
Moreover, when $N_\textsc{ue} = 8$, the channel capacity is approximately twice as large as when $N_\textsc{ue} = 4$.
This means that, when $M \gg N$, deploying more antennas on a single UT can proportionally increase the throughput per UT at the cost of weakening the channel hardening.

{\em Case Study 5:}
This case study aims to showcase the parameter fitting results of ELAA-MIMO channel capacity in various typical scenarios.
The system variable parameters are configured by: $d_\perp = 1$, $25$, $50$ m, $N_{\textsc{ue}} = 4$, $8$ and $M = 2,000$ and $200,000$, for the three cases of the proposed channel model ($36$ scenarios in total).
Moreover, the parametric fitting is carried out at $R = 11$ different average-SNRs, which are sampled uniformly from $10$ to $30$ $\mathrm{dB}$.
The results of parametric fitting in two of the scenarios are illustrated in \figref{fig06}.
It can be found that the CDF of ELAA-MIMO channel capacity can be well fitted by the skew normal distribution in both of the pictures.
Moreover, \figref{fig06a} shows that the channel capacity is close to the Gaussian distribution when $d_{\perp} = 1$ m; 
\figref{fig06b} shows that the channel capacity is close to the Weibull distribution when $d_{\perp} = 50$ m.

\begin{figure*}[t]
	\centering
	\subfigure[$\widehat{\theta}_1$ of the skew normal distribution]{
		\centering
		\includegraphics[width=0.45\textwidth]{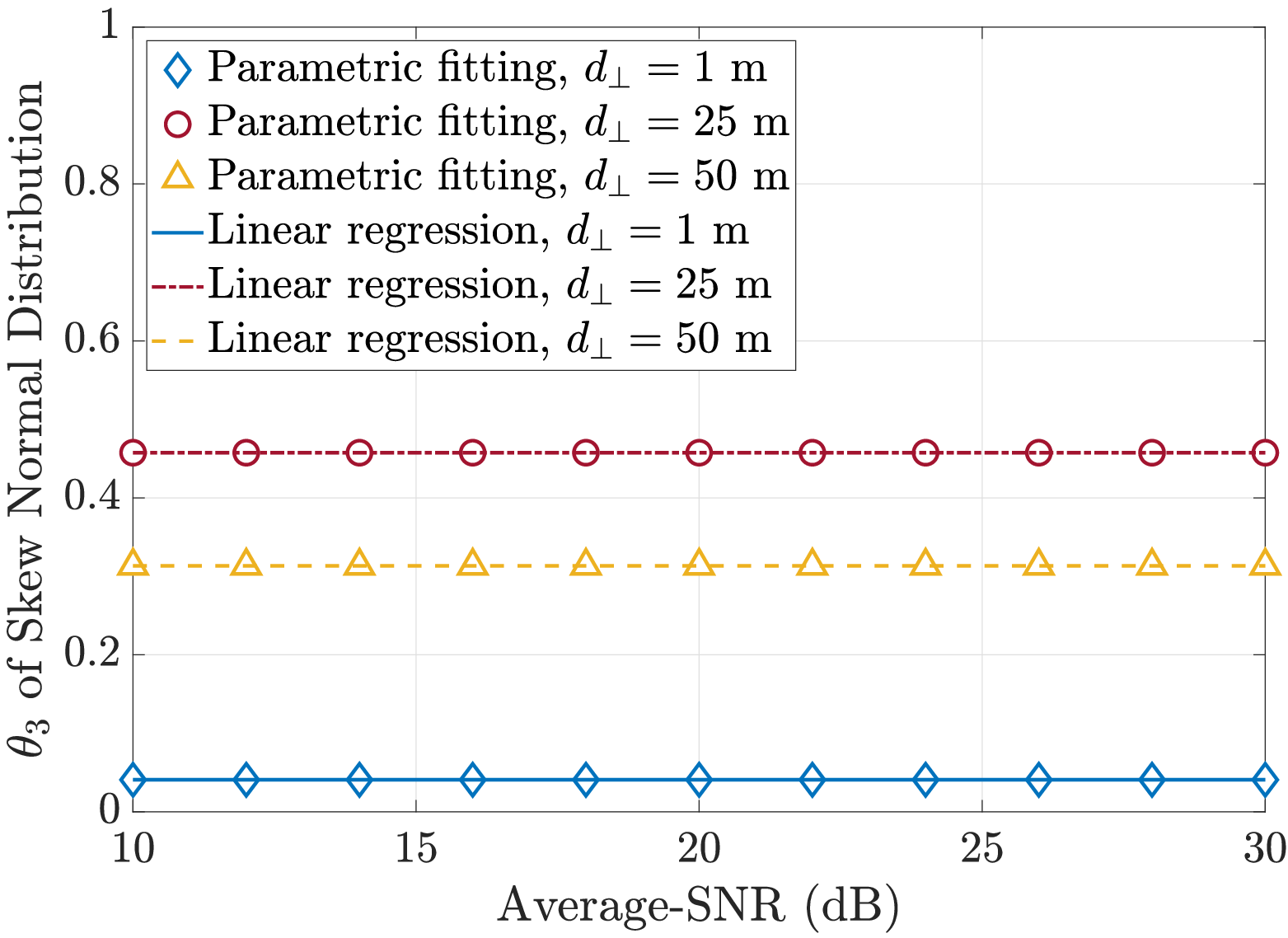}}
	\subfigure[$\widehat{\theta}_2$ of the Weibull distribution]{
		\centering
		\includegraphics[width=0.45\textwidth]{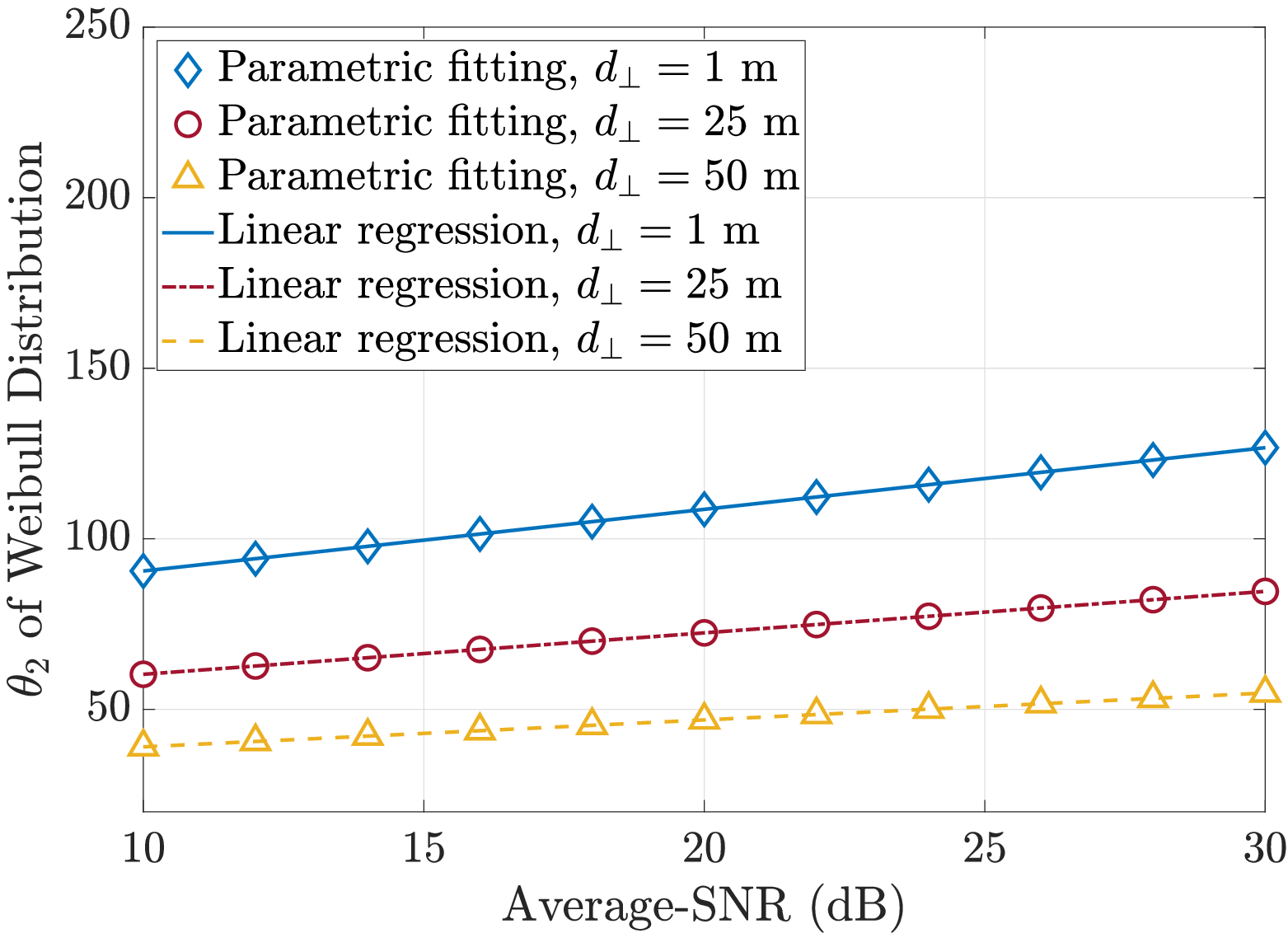}}
	\caption{\label{fig07}Two examples of parametric fitting and linear regression results. {\bf Case 2}; $N_{\textsc{ue}} = 8$.}
	\vspace{-0em}
\end{figure*}
The values of estimated parameters when $N_{\textsc{ue}} = 4$ at $\gamma_o = 10$ $\mathrm{dB}$ are shown in \appref{appD}.
As shown in \tabref{tab01} ($M = 2,000$) and \tabref{tab02} ($M = 200,000$), the skew normal distribution has the lowest fitting error in all the scenarios.
Moreover, it can also be found that, for {\bf Case 3} and $d_{\perp} = 1$ m, the channel capacity is close to the Gaussian distribution; for the multiuser scenarios with relatively large $d_\perp$, the channel capacity is close to the Weibull distribution.
Due to the page limitation, the tables of parametric fitting results when $N_\textsc{ue} = 8$ at other average-SNRs are uploaded onto GitHub at the following link: \url{https://github.com/ELAA-MIMO/non-stationary-fading-channel-model}.

{\em Case Study 6:}
\begin{figure*}[t]
	\centering
	\includegraphics[width=0.93\textwidth]{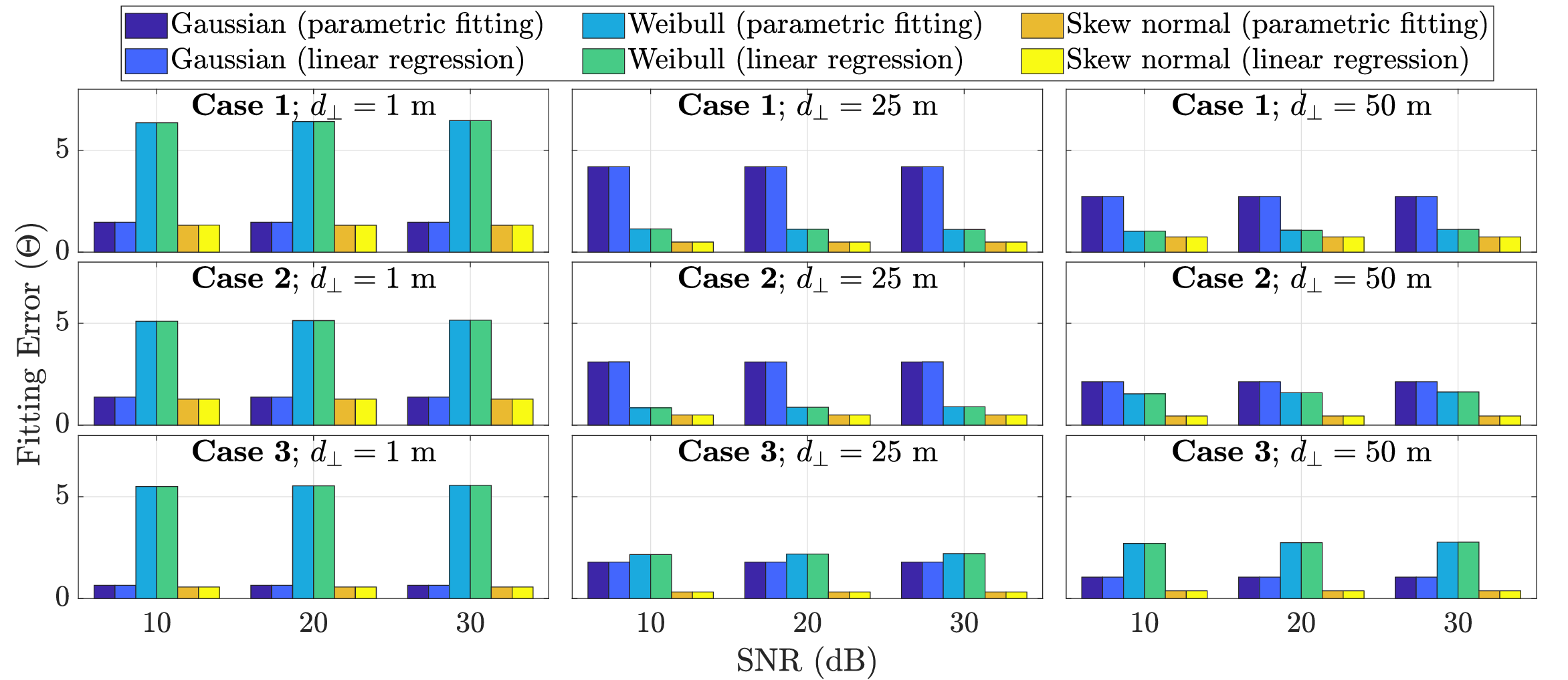}
	\vspace{-0.5em}
	\caption{\label{fig08}The fitting errors of parametric fitting and linear regression when $N_\textsc{ue} = 4$ and $M = 200,000$.}
	\vspace{-1em}
\end{figure*}
\begin{figure*}[t]
	\centering
	\subfigure[]{
		\centering
		\includegraphics[width=0.45\textwidth]{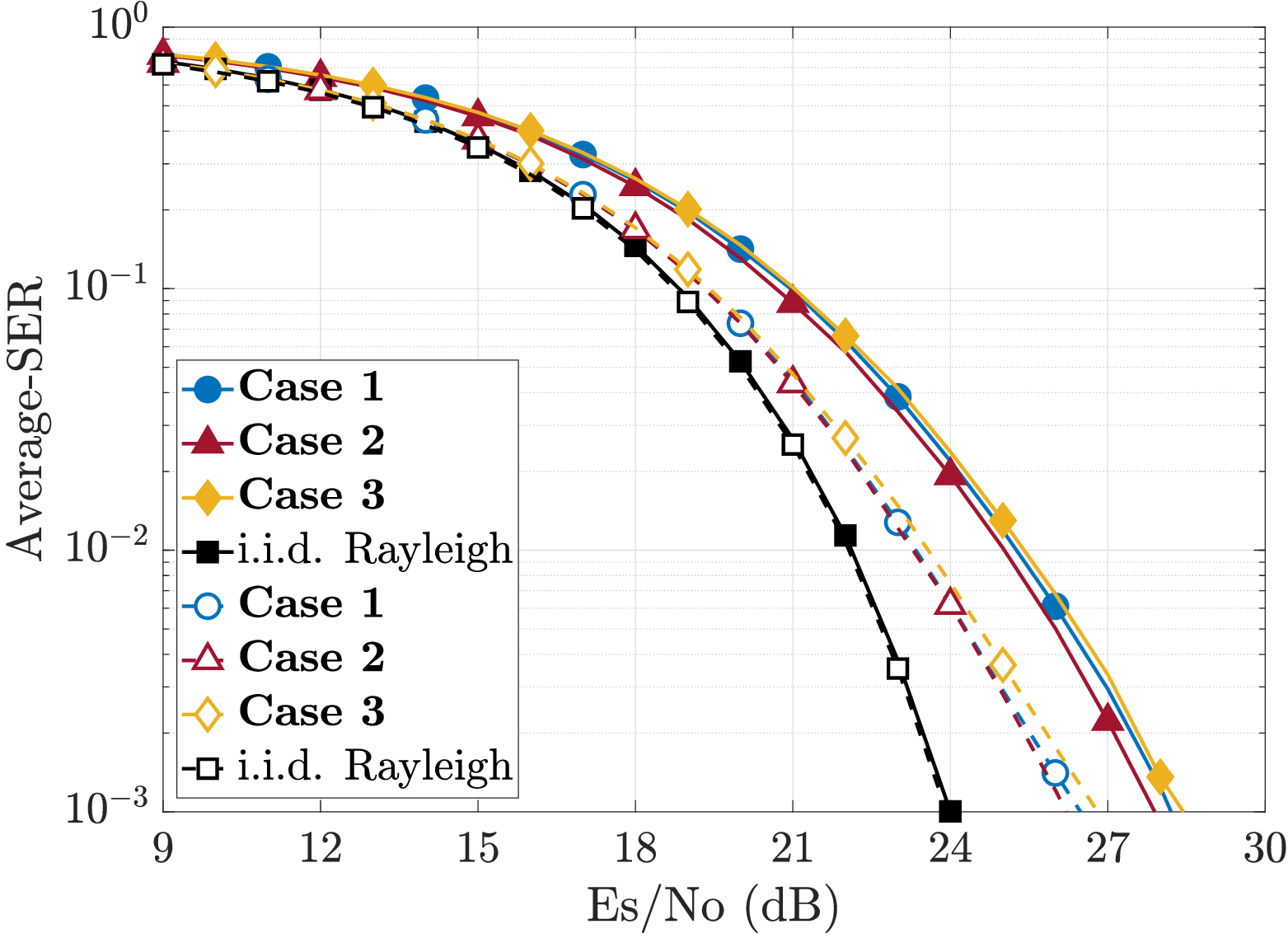}
		\label{fig09a}}	
	\subfigure[]{
		\centering
		\includegraphics[width=0.45\textwidth]{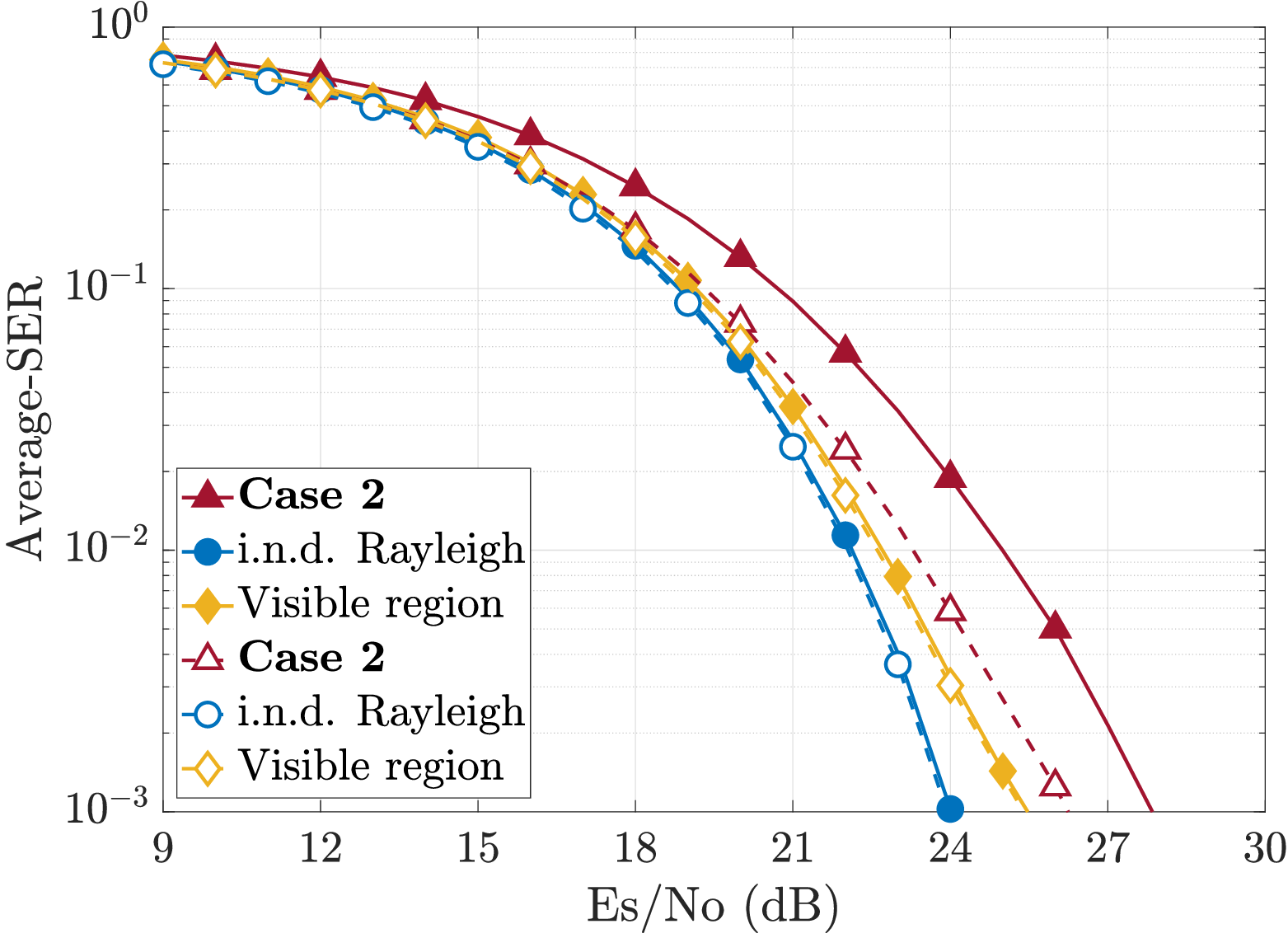}
		\label{fig09b}	}
	\caption{\label{fig09} The performance of LMMSE receiver in different types of MIMO channels when $M = 2000$, $N = 20$, and $N_\textsc{ue} = 4$. The modulation is $64$ QAM.
		Solid lines: LMMSE;
		dashed lines: MRC lower-bound.}
	\vspace{-0.5em}
\end{figure*}
This case study aims to showcase the relationship between $\gamma_o$ and each of the estimated parameters.
The results of linear regression for the two estimated parameters are illustrated in \figref{fig07}: $\widehat{\theta}_1$ of the skew normal distribution, and $\widehat{\theta}_2$ of the Weibull distribution.
As shown in this figure, the relationship between $\gamma_o$ and each of the two parameters is almost linear.
Based on our investigations, the relationship between $\gamma_o$ and each of the other estimated parameters is also approximately linear.
Moreover, when $N_\textsc{ue} = 4$, the values of $a$ and $c$ obtained by linear regression are shown in \appref{appD}.
This results can be used to analysis the channel capacity directly, avoiding the generation of ELAA-MIMO channel if not necessary.
All the other results of linear regression are uploaded onto GitHub as well.

The errors ($\Theta$) of parametric fitting and linear regression are shown in \figref{fig08}, where $N_\textsc{ue} = 4$ and $M = 200,000$.
As shown in the figure, the values of $\Theta$ are almost the same for parametric fitting and linear regression.
Such the results confirm that each of the estimated parameters has an approximately linear relationship with $\gamma_o$.
Again, the ELAA-MIMO channel capacity obeys the skew normal distribution, and for the Gaussian and Weibull distributions, their suitable scenarios are the same as discussed in {\em Case Study 5}.
Moreover, the same conclusion can also be drawn from the numerical results in the scenarios when $N_{\textsc{ue}} = 8$ and $M = 2,000$ as well. 
Again, due to the page limitation, these results are uploaded onto GitHub.

\section{Evaluation of Linear Receivers in spatially Non-stationary ELAA-MIMO Channels}\label{sec4}
The objective of this section is to investigate impact of the spatial non-stationarity on linear MIMO receivers.
Taking \gls{lmmse} as an example, the estimated symbol of $\mathbf{s}$, denoted by $\hat{\mathbf{s}}$, is as follows \cite{Li2022}
\begin{equation}
	\hat{\mathbf{s}} = \Big(\mathbf{H}^H\mathbf{H} + \frac{\gamma_o}{N}\mathbf{I}\Big)^{-1}\mathbf{H}^H\mathbf{y}.
\end{equation}
In this section, in addition to the proposed channel model, the following two spatially non-stationary channel models are also investigated:

{\em 1)} i.n.d. Rayleigh: the spatially non-stationary Rayleigh fading channel in \cite{Amiri2018}.
This is a special case of the proposed channel model by setting $p_\textsc{los}(d^\mathrm{2D}_{m,n}) = 0, \forall m, n $ and $\epsilon_{m,n} = 0, \forall m, n$;

{\em 2)} Visible Region: the spatially non-stationary channel model proposed in \cite{Rodrigues2020}.
The visibility region is randomly distributed on the ULA with its length following the log-normal distribution ($L_{\textsc{vr}} \sim \mathcal{LN}(\ln5, \ln0.4)$ in this case study). 
The channel is generated according to \eqref{eqn07}.

The statistical behaviors of the above spatially non-stationary channels have already been discussed in our previous work (see \cite{Liu2021}).
Also, it is not our aim here to compare the different channel models in terms of their accuracy. 
The accuracy and practicality of the proposed channel model have been well reflected in \secref{sec2d}.

In this section, the propagation environment is set the same as {\em Case Study 1}.
The ULA is deployed with $M = 2,000$ antennas and the UT-to-ELAA distance is set as $d_\perp = 50$ m.
There are $5$ UTs, each with $N_{\textsc{ue}} = 4$ antennas, deployed in parallel to the ELAA.
Since $N = 20$ and the modulation is $64$ QAM, it is computational prohibitive to perform \gls{mlsd} by Monte Carlo simulation. 
As an instead, the performance baseline is set as \gls{mrc} lower-bound, where the $n^{th}$ estimated symbol is defined as follows \cite{Amiri2022}
\begin{equation}
	\hat{s}_n = \mathbf{h}^{H}_{n}\mathbf{h}_{n} s_{n} + \mathbf{h}^{H}_{n}\mathbf{v},
\end{equation}
which means that the interference is perfectly removed and then MRC is used for each interference-free data stream.

The performance of LMMSE receiver in different types of MIMO channels is shown in \figref{fig09}.
In \figref{fig09a}, the LMMSE receiver has a near-optimal performance in the i.i.d. Rayleigh fading channel.
However, in the three cases of proposed channel model, the performance gaps between the LMMSE receiver and MRC bound is significant (about $2$ $\mathrm{dB}$).
In \figref{fig09b}, the LMMSE receiver also has near-optimal performances in both i.n.d. Rayleigh fading channel and visible region model.
The reason is that both of the two spatially non-stationary channel models have their channel non-stationarity mainly coming from the spherical-wave propagation, but their small-scale fading is still WSSUS.
For the same reason, it can be found that the performance of LMMSE is approximately the same in the i.i.d. and i.n.d. Rayleigh channels.
The performance of LMMSE in the visible region model is worse than that in the i.n.d. Rayleigh channel. 
This is because the transmitted power in the invisible region is considered to be lost.

\section{Conclusion and Outlook}\label{sec5}
In this paper, a novel spatially non-stationary channel model has been proposed for link-level study of ELAA-MIMO systems.
Unlike current spatially non-stationary channel models, the correlation of LoS/NLoS states between different UTs are considered in the proposed channel model and the correlation is modeled by S-OBE method.
The pseudocode, which is simple and easy to implement on the link-level, has been provided for Monte Carlo simulations.
It has been demonstrated that the proposed channel model can capture almost all the spatially non-stationary properties, including spherical wavefront, shadowing effects, correlations of LoS states, and many physical differences between the LoS and NLoS links.
Moreover, it has been demonstrated that the channel data generated by the proposed channel model can well match the measured data from practical ELAA channels.

It has been shown that the channel spatial non-stationarity is detrimental to the MIMO channel hardening.
The variance of ELAA-MIMO channel capacity is considerably large, even when the number of service-antennas is extremely large.
Moreover, it has been found that both inter-UT and intra-UT correlations render the ELAA-MIMO channel more ill-conditioned.
The distribution of ELAA-MIMO channel capacity is then studied by parametric fitting and linear regression.
In all of our study scenarios, the channel capacity has been found to obey the skew normal distribution.
Moreover, the channel capacity has also been found close to the Gaussian or Weibull distribution.
More specifically, for the single-user equivalent scenarios or multiuser scenarios with short UT-to-ELAA distances, the channel capacity is close to the Gaussian distribution;
and for multiuser scenarios with relative large UT-to-ELAA distances, the channel capacity is close to the Weibull distribution.

In addition, it has been demonstrated that current linear MIMO receivers cannot offer near-optimum detection performance in spatially non-stationary channels.
A fundamental redesign of MIMO receivers (and perhaps also transmitters) is therefore required for future ELAA-MIMO technology. 
Potential research directions include uplink power control, spatial non-stationarity incorporated linear receiver (or precoding) architectures, 
as well as their optimization in the context of capacity-constrained fronthaul/backhaul systems.
Finally, the focus of this paper is on the development of a narrowband fading channel model as this allowed for a more simplified and tractable analysis.
The proposed model can be extended to the wideband case by incorporating additional parameters that capture the spectral characteristics of the channel.
This could be the further research direction of this paper that exploring the performance of wideband ELAA-MIMO systems.

\appendices
\section{Proof of Corollary \ref{cor1}}\label{appA}
Consider the case when there only have $L$ adjacent service-antennas sharing the same LoS (or NLoS) state. 
Those antennas are labelled by an index set $\{\ell, \ell+1, ...,\ell+L-1\}$.
Service-antennas outside this set have their LoS/NLoS states uncorrelated with those within the set. 
Given $L$ a random variable, the PMF of $L$ is derived as
\begin{IEEEeqnarray}{ll}
	f_{L}(x) &= \left(1 - p_{\textsc{los}}(\ell,\ell + x )\right) \prod_{m = \ell + 1}^{\ell + x -1 }p_{\textsc{los}}(\ell,m) \nonumber \\
	&= \prod_{m = \ell + 1}^{\ell + x -1 }p_{\textsc{los}}(\ell,m) -  \prod_{m = \ell + 1}^{\ell + x }p_{\textsc{los}}(\ell,m). \label{eqn55}
\end{IEEEeqnarray}
Let $\ell = 0$ for the sake of notation simplification. 
We plug \eqref{eqn15} into \eqref{eqn55} and obtain
\begin{IEEEeqnarray}{ll}
	f_{L}(x) &\nonumber=\prod_{m=1}^{x-1}\exp\Big(-\dfrac{\lambda m}{2d_\textsc{los}}\Big)-\prod_{m=1}^{x}\exp\Big(-\dfrac{\lambda m}{2d_\textsc{los}}\Big) \\
	&\nonumber=\exp\Big(\frac{\lambda \sum_{m=1}^{x-1}m}{-2d_\textsc{los}}\Big) - \exp\Big(\frac{\lambda  \sum_{m=1}^{x}m}{-2d_\textsc{los}}\Big) \\
	&\label{eqn56}= \exp\Big(\dfrac{\lambda(x^2 -x)}{-4d_{\textsc{los}}}\Big) - \exp\Big(\dfrac{\lambda(x^2 + x)}{-4d_{\textsc{los}}}\Big).
\end{IEEEeqnarray}
{\em Corollary \ref{cor1}} is therefore proved.

\section{Pseudocode Used to Generate $\mat{\beta}$ for RUs}\label{appB}
\begin{algorithm}[H]
	{{\small \caption{{Random realization of} $\mat{\beta}$ for RUs} 
			\begin{algorithmic}[1]\label{alg01}
				\renewcommand{\algorithmicrequire}{\textbf{Input:}} 
				\REQUIRE~\\ $M$: the number of service-antennas;\\ $d_\ell^{\textsc{2d}}$: the 2-D distance used in \eqref{eqn12};\\ 
				$\lambda, d_\textsc{los}$: parameters used in \eqref{eqn14};
				\renewcommand{\algorithmicrequire}{\textbf{Output:}} 
				\REQUIRE~\\${\mat{\beta}}$: the LoS/NLoS state vector;
				\renewcommand{\algorithmicensure}{\textbf{START}}
				\ENSURE  
				\STATE {\bf let} $\ell=0$; call \eqref{eqn12} to compute $p_{\textsc{los}}(d_\ell^{\textsc{2d}})$;
				\STATE {\bf let} $\varrho_\ell=p_{\textsc{los}}(d_\ell^{\textsc{2d}})$ and generate $\beta_\ell$ according to the Bernoulli distribution in \eqref{eqn10}; {\bf let} $m=\ell+1$;
				\STATE {Generate} $\beta_m$ according to the probability in \eqref{eqn11};
				\STATE {\bf if} ${\beta}_m={\beta}_\ell$, {\bf then} $m\leftarrow m+1$; {\bf otherwise} $\ell\leftarrow m$, {\bf goto} step 2; 
				\STATE {\bf repeat} step 3 until $m=M$; 
				\renewcommand{\algorithmicensure}{\textbf{END}}
				\ENSURE
	\end{algorithmic} }}
\end{algorithm}

It is perhaps worth noting on Step $4$:  for $\beta_m\neq \beta_\ell$, this is the case when $\beta_m$ should be considered as uncorrelated with $\beta_\ell$. 
Then, $\beta_m$ must be independently generated by going back to Step $2$.

\section{Pseudocode Used to Generate $\mat{\beta}_{k}$ for non-RUs}\label{appC}
\begin{algorithm}[http]
	{\small \caption{{Random realization of} $\mat{\beta}$ for non-RUs in the linear Wyner-type model} 
		\begin{algorithmic}[1]\label{alg02}
			\renewcommand{\algorithmicrequire}{\textbf{Input:}} 
			\REQUIRE~\\$\mat{\beta}_j, \mat{\beta}_{j+1}$: LoS/NLoS states of the corresponding RU places (the output of \algref{alg01});
			\\$d_{j,k}$, $d_{j+1,k}$: the corresponding distances in \eqref{eqn31};
			\renewcommand{\algorithmicrequire}{\textbf{Output:}} 
			\REQUIRE~\\$\mat{\beta}_k$: LoS state of the $k^{th}$ non-RU;
			\renewcommand{\algorithmicensure}{\textbf{START}}
			\ENSURE  
			\STATE Compute $\zeta_{k,j}, \zeta_{k,j+1}$ according to \eqref{eqn31};
			\STATE Compute $\mat{\varrho}_k$ according to \eqref{eqn30}; 
			\STATE Use $\mat{\varrho}_k$ to generate $\mat{\beta}_k$ using the same procedure as in {\bf Algorithm \ref{alg01}};
			\renewcommand{\algorithmicensure}{\textbf{END}}
			\ENSURE
	\end{algorithmic} }
\end{algorithm}
\vspace{-1em}

\section{The List of Statistical Models} \label{appR1}
The following statistical models are attempted by parametric fitting in this paper: Bernoulli, Beta, Binomial, Birnbaum-Saunders, Burr, exponential, Gamma, Gaussian, geometric, generalized extreme value, generalized Pareto, half-normal, inverse Gaussian, log-logistic, logistic, log normal, Nakagami, negative Binomial, Poisson, Rayleigh, Rician, skew normal, stable, t location-scale, uniform, and Weibull distributions.

\section{Results of Parametric Fitting and Linear Regression}\label{appD}
Here are results of parametric fitting ($\gamma_o = 10$ dB) and linear regression when $N_{\textsc{ue}} = 4$.
All the other results are uploaded to GitHub due to the page limitation.
See \url{https://github.com/ELAA-MIMO/Non-stationary-fading-channel-model/} for more results.

\begin{table*}[http!]
	\renewcommand{\arraystretch}{1.18}
	\caption{The results of parametric fitting when $N_{\textsc{ue}} = 4$, $\gamma_o = 10\ \mathrm{dB}$, and $M = 2,000$.}
	\vspace{-0em}
	\label{tab01}
	\centering
	\resizebox{1\textwidth}{!}{
		\begin{tabular}{|c|c|c|c|c|c|c|c|c|c|c|}
			\hline
			\multirow{2}{*}{}  & {Case} & \multicolumn{3}{c|}{\bf Case 1}          
			& \multicolumn{3}{c|}{\bf Case 2}  & \multicolumn{3}{c|}{\bf Case 3}  \\ \cline{2-11} 
			& $d_\perp$  & $1$ m  & $25$ m    & $50$ m   & $1$ m   & $25$ m  & $50$ m & $1$ m& $25$ m &$50$ m    \\ \hline
			\multirow{2}{*}{\begin{tabular}[c]{@{}c@{}}Monte Carlo\\ Trial \end{tabular}}
			& $\mu_T$ & $196.6$ & $194.5$ & $185.7$ & $196.5$ & $192.7$ & $184.9$ & $196.7$ & $192.5$ & $184.2$\\\cline{2-11}
			& $\sigma_T$ & $3.513$ & $9.799$ & $16.62$ & $3.521$ & $7.721$ & $12.51$ & $2.943$ & $5.172$ & $8.096$\\ \hline
			\multirow{3}{*}{\begin{tabular}[c]{@{}c@{}}Gaussian \\ Distribution\end{tabular}}
			& $\theta_1$ & $196.6$ & $194.5$ & $185.7$ & $196.5$ & $192.7$ & $184.9$ & $196.7$ & $192.5$ & $184.2$\\\cline{2-11}
			& $\theta_2$ & $3.513$ & $9.799$ & $16.62$ & $3.521$ & $7.721$ & $12.51$ & $2.943$ & $5.172$ & $8.096$\\\cline{2-11}
			& $\Theta$ & $1.421$ & $4.542$ & $3.573$ & $1.195$ & $3.643$ & $2.441$ & $0.6854$ & $2.186$ & $1.217$\\ \hline
			\multirow{3}{*}{\begin{tabular}[c]{@{}c@{}}WeiBull\\ Distribution \end{tabular}}
			& $\theta_1$ & $198.3$ & $198.8$ & $193.0$ & $198.2$ & $196.2$ & $190.6$ & $198.2$ & $195.0$ & $188.0$\\\cline{2-11}
			& $\theta_2$ & $54.54$ & $26.00$ & $13.88$ & $49.93$ & $30.92$ & $17.50$ & $63.34$ & $42.76$ & $25.23$\\\cline{2-11}
			& $\Theta $ & $4.491$ & $2.143$ & $1.219$ & $5.701$ & $0.8589$ & $1.003$ & $4.942$ & $1.474$ & $2.213$\\ \hline
			\multirow{4}{*}{\begin{tabular}[c]{@{}c@{}}Skew\\ Normal \\ Distribution\end{tabular}}
			& $\theta_1$ & $197.0$ & $204.9$ & $200.1$ & $196.5$ & $198.9$ & $192.5$ & $196.5$ & $195.1$ & $186.6$\\\cline{2-11}
			& $\theta_2$ & $3.508$ & $8.728$ & $15.68$ & $3.520$ & $7.282$ & $12.18$ & $2.942$ & $5.063$ & $8.037$\\\cline{2-11}
			& $\theta_3$ & $0.07264$ & $0.7515$ & $0.5681$ & $0.01651$ & $0.5295$ & $0.3814$ & $-0.03665$ & $0.3191$ & $0.1868$\\ \cline{2-11}
			& $\Theta$ & $1.178$ & $0.3798$ & $0.8131$ & $1.161$ & $0.4276$ & $0.6938$ & $0.6135$ & $0.3524$ & $0.394$\\ \hline
	\end{tabular}}
	\vspace{-0em}
\end{table*}

\begin{table*}[http!]
	\renewcommand{\arraystretch}{1.18}
	\caption{The results of parametric fitting when $N_{\textsc{ue}} = 4$, $\gamma_o = 10\ \mathrm{dB}$, and $M = 200,000$.}
	\vspace{-0em}
	\label{tab02}
	\centering
	\resizebox{1\textwidth}{!}{
		\begin{tabular}{|c|c|c|c|c|c|c|c|c|c|c|}
			\hline
			\multirow{2}{*}{}  & {Case} & \multicolumn{3}{c|}{\bf Case 1}          
			& \multicolumn{3}{c|}{\bf Case 2}  & \multicolumn{3}{c|}{\bf Case 3}  \\ \cline{2-11} 
			& $d_\perp$ & $1$ m & $25$ m & $50$ m & $1$ m & $25$ m & $50$ m & $1$ m& $25$ m &$50$ m  \\ \hline
			\multirow{2}{*}{\begin{tabular}[c]{@{}c@{}}Monte Carlo\\ Trial \end{tabular}}
			& $\mu_T$ & $329.2$ & $328.8$ & $325.1$ & $329.0$ & $328.0$ & $325.2$ & $329.4$ & $327.9$ & $324.6$\\\cline{2-11}
			& $\sigma_T$ & $3.536$ & $8.575$ & $12.82$ & $3.435$ & $6.573$ & $9.594$ & $2.837$ & $4.447$ & $6.184$\\ \hline
			\multirow{3}{*}{\begin{tabular}[c]{@{}c@{}}Gaussian \\ Distribution\end{tabular}}
			& $\theta_1$ & $329.2$ & $328.8$ & $325.1$ & $329.0$ & $328.0$ & $325.2$ & $329.4$ & $327.9$ & $324.6$\\\cline{2-11}
			& $\theta_2$ & $3.536$ & $8.575$ & $12.82$ & $3.435$ & $6.573$ & $9.594$ & $2.837$ & $4.447$ & $6.184$\\\cline{2-11}
			& $\Theta$ & $1.471$ & $4.194$ & $2.735$ & $1.375$ & $3.101$ & $2.128$ & $0.6466$ & $1.791$ & $1.049$\\ \hline
			\multirow{3}{*}{\begin{tabular}[c]{@{}c@{}}WeiBull\\ Distribution \end{tabular}}
			& $\theta_1$ & $330.9$ & $332.7$ & $331.0$ & $330.7$ & $331.1$ & $329.7$ & $330.8$ & $330.0$ & $327.6$\\\cline{2-11}
			& $\theta_2$ & $79.46$ & $48.87$ & $30.29$ & $90.57$ & $60.29$ & $39.03$ & $103.9$ & $81.92$ & $56.80$\\\cline{2-11}
			& $\Theta $ & $6.349$ & $1.143$ & $1.027$ & $5.096$ & $0.8505$ & $1.534$ & $5.500$ & $2.157$ & $2.703$\\ \hline
			\multirow{4}{*}{\begin{tabular}[c]{@{}c@{}}Skew\\ Normal \\ Distribution\end{tabular}}
			& $\theta_1$ & $329.4$ & $336.5$ & $333.8$ & $329.3$ & $332.6$ & $329.9$ & $329.2$ & $329.6$ & $326.2$\\\cline{2-11}
			& $\theta_2$ & $3.534$ & $7.911$ & $12.37$ & $3.433$ & $6.285$ & $9.399$ & $2.835$ & $4.386$ & $6.151$\\\cline{2-11}
			& $\theta_3$ & $0.04767$ & $0.6089$ & $0.4326$ & $0.04075$ & $0.4575$ & $0.3134$ & $-0.0442$ & $0.2435$ & $0.1590$\\ \cline{2-11}
			& $\Theta$ & $1.327$ & $0.497$ & $0.7451$ & $1.279$ & $0.4995$ & $0.4464$ & $0.5646$ & $0.3207$ & $0.3773$\\ \hline
	\end{tabular}}
	\vspace{-0em}
\end{table*}

\begin{table*}[http!]
	\renewcommand{\arraystretch}{1.18}
	\caption{The results of linear regression when $N_{\textsc{ue}} = 4$, and $M = 2,000$.}	\vspace{-0em}
	\label{tab03}
	\centering
	\resizebox{1\textwidth}{!}{
		\begin{tabular}{|c|c|c|c|c|c|c|c|c|c|c|c|}
			\hline
			\multicolumn{2}{|c|}{\multirow{2}{*}{}}   & {Case}  & \multicolumn{3}{c|}{\bf Case 1}   & \multicolumn{3}{c|}{\bf Case 2}   & \multicolumn{3}{c|}{\bf Case 3} \\ \cline{3-12} 			
			\multicolumn{2}{|c|}{} & $d_\perp$  & $1$ m  & $25$ m & $50$ m & $1$ m  & $25$ m & $50$ m & $1$ m  & $25$ m & $50$ m  \\ \hline			
			\multirow{4}{*}{\begin{tabular}[c]{@{}c@{}}Gaussian\\ Distribution\end{tabular}}  & \multirow{2}{*}{$\theta_1$}
			& $a$ & $6.642$ & $6.642$ & $6.641$ & $6.642$ & $6.642$ & $6.641$ & $6.642$ & $6.642$ & $6.641$\\ \cline{3-12}
			& & $c$ & $130.2$ & $128.1$ & $119.3$ & $130.0$ & $126.3$ & $118.5$ & $130.3$ & $126.1$ & $117.7$\\ \cline{2-12}
			& \multirow{2}{*}{$\theta_2$}
			& $a \times 10^{6}$ & $192.5$ & $809.7$ & $1726$ & $187.7$ & $660.7$ & $1310$ & $149.2$ & $469.7$ & $924.0$\\ \cline{3-12}
			& & $c$ & $3.513$ & $9.797$ & $16.61$ & $3.521$ & $7.720$ & $12.51$ & $2.943$ & $5.172$ & $8.095$\\ \hline
			\multirow{4}{*}{\begin{tabular}[c]{@{}c@{}}Weibull\\ Distribution\end{tabular}} & \multirow{2}{*}{$\theta_1$}
			& $a$ & $6.643$ & $6.646$ & $6.652$ & $6.643$ & $6.644$ & $6.648$ & $6.643$ & $6.643$ & $6.644$\\ \cline{3-12}
			& & $c$ & $131.9$ & $132.3$ & $126.4$ & $131.7$ & $129.7$ & $124.1$ & $131.7$ & $128.5$ & $121.5$\\ \cline{2-12}
			& \multirow{2}{*}{$\theta_2$}    & $a$ & $1.804$ & $0.8830$ & $0.4899$ & $1.637$ & $1.054$ & $0.6158$ & $2.100$ & $1.454$ & $0.8882$\\ \cline{3-12}
			& & $c$ & $36.49$ & $17.16$ & $8.972$ & $33.55$ & $20.36$ & $11.33$ & $42.32$ & $28.19$ & $16.33$\\ \hline
			\multirow{6}{*}{\begin{tabular}[c]{@{}c@{}}Skew \\ Normal\\ Distribution\end{tabular}}    & \multirow{2}{*}{$\theta_1$}
			& $a$ & $6.643$ & $6.643$ & $6.643$ & $6.643$ & $6.643$ & $6.642$ & $6.643$ & $6.642$ & $6.641$\\ \cline{3-12}
			& & $c$ & $130.6$ & $138.5$ & $133.7$ & $130.1$ & $132.4$ & $126.0$ & $130.1$ & $128.7$ & $120.2$\\ \cline{2-12}
			& \multirow{2}{*}{$\theta_2$} & $a \times 10^{7}$ & $204.0$ & $172.5$ & $495.5$ & $217.7$ & $255.0$ & $470.2$ & $161.6$ & $266.1$ & $382.5$\\ \cline{3-12}
			& & $c \times 10^{2}$ & $7.261$ & $75.15$ & $56.80$ & $1.647$ & $52.95$ & $38.14$ & $-3.667$ & $31.91$ & $18.67$\\ \cline{2-12}
			& \multirow{2}{*}{$\theta_3$} & $a \times 10^{6}$ & $189.0$ & $642.0$ & $1429$ & $186.9$ & $576.5$ & $1186$ & $150.1$ & $440.8$ & $892.8$\\ \cline{3-12}
			& & $c$ & $3.507$ & $8.727$ & $15.67$ & $3.520$ & $7.281$ & $12.17$ & $2.942$ & $5.062$ & $8.036$\\ \hline
	\end{tabular}}
	\vspace{-0em}
\end{table*}

\begin{table*}[http!]
	\renewcommand{\arraystretch}{1.18}
	\caption{The results of linear regression when $N_{\textsc{ue}} = 4$, and $M = 200,000$.}
	\vspace{-0em}
	\label{tab04}
	\centering
	\resizebox{1\textwidth}{!}{
		\begin{tabular}{|c|c|c|c|c|c|c|c|c|c|c|c|}
			\hline
			\multicolumn{2}{|c|}{\multirow{2}{*}{}}   & {Case}  & \multicolumn{3}{c|}{\bf Case 1}   & \multicolumn{3}{c|}{\bf Case 2}   & \multicolumn{3}{c|}{\bf Case 3} \\ \cline{3-12} 			
			\multicolumn{2}{|c|}{} & $d_\perp$  & $1$ m  & $25$ m & $50$ m & $1$ m  & $25$ m & $50$ m & $1$ m  & $25$ m & $50$ m  \\ \hline			
			\multirow{4}{*}{\begin{tabular}[c]{@{}c@{}}Gaussian\\ Distribution\end{tabular}}  & \multirow{2}{*}{$\theta_1$}
			& $a$ & $6.644$ & $6.644$ & $6.644$ & $6.644$ & $6.644$ & $6.644$ & $6.644$ & $6.644$ & $6.644$\\ \cline{3-12}
			& & $c$ & $262.7$ & $262.4$ & $258.7$ & $262.6$ & $261.6$ & $258.8$ & $262.9$ & $261.4$ & $258.2$\\ \cline{2-12}
			& \multirow{2}{*}{$\theta_2$}
			& $a \times 10^{6}$ & $1.922$ & $6.405$ & $10.57$ & $1.875$ & $4.999$ & $7.829$ & $1.472$ & $3.512$ & $5.286$\\ \cline{3-12}
			& & $c$ & $3.536$ & $8.575$ & $12.82$ & $3.435$ & $6.573$ & $9.594$ & $2.837$ & $4.447$ & $6.184$\\ \hline
			\multirow{4}{*}{\begin{tabular}[c]{@{}c@{}}Weibull\\ Distribution\end{tabular}} & \multirow{2}{*}{$\theta_1$}
			& $a$ & $6.644$ & $6.645$ & $6.647$ & $6.644$ & $6.645$ & $6.645$ & $6.644$ & $6.644$ & $6.645$\\ \cline{3-12}
			& & $c$ & $264.4$ & $266.2$ & $264.6$ & $264.3$ & $264.6$ & $263.2$ & $264.3$ & $263.6$ & $261.1$\\ \cline{2-12}
			& \multirow{2}{*}{$\theta_2$}    & $a$ & $1.564$ & $0.9838$ & $0.613$ & $1.807$ & $1.215$ & $0.7892$ & $2.068$ & $1.649$ & $1.151$\\ \cline{3-12}
			& & $c$ & $63.83$ & $39.03$ & $24.16$ & $72.5$ & $48.15$ & $31.14$ & $83.22$ & $65.43$ & $45.29$\\ \hline
			\multirow{6}{*}{\begin{tabular}[c]{@{}c@{}}Skew \\ Normal\\ Distribution\end{tabular}}    & \multirow{2}{*}{$\theta_1$}
			& $a$ & $6.644$ & $6.644$ & $6.644$ & $6.644$ & $6.644$ & $6.644$ & $6.644$ & $6.644$ & $6.644$\\ \cline{3-12}
			& & $c$ & $263.0$ & $270.1$ & $267.3$ & $262.8$ & $266.2$ & $263.5$ & $262.7$ & $263.1$ & $259.7$\\ \cline{2-12}
			& \multirow{2}{*}{$\theta_2$} & $a \times 10^{7}$ & $1.979$ & $2.240$ & $4.250$ & $2.118$ & $2.554$ & $3.406$ & $1.552$ & $2.261$ & $2.516$\\ \cline{3-12}
			& & $c \times 10^{2}$ & $4.767$ & $60.89$ & $43.26$ & $4.075$ & $45.75$ & $31.34$ & $-4.420$ & $24.35$ & $15.90$\\ \cline{2-12}
			& \multirow{2}{*}{$\theta_3$} & $a \times 10^{6}$ & $1.902$ & $5.341$ & $9.207$ & $1.848$ & $4.447$ & $7.211$ & $1.489$ & $3.370$ & $5.161$\\ \cline{3-12}
			& & $c$ & $3.534$ & $7.911$ & $12.37$ & $3.433$ & $6.285$ & $9.398$ & $2.835$ & $4.386$ & $6.151$\\ \hline
	\end{tabular}}
	\vspace{-0em}
\end{table*}

\bibliographystyle{IEEEtran}
\bibliography{../IEEEabrv,../mMIMO,../ymbib} 
	
\end{document}